\documentclass[useAMS,usegraphicx,usenatbib]{mn2e}
\usepackage{amssymb}

\newcommand{\msun}{\mbox{M$_\odot$}}% Msun
\title[Massive jets from intermediate-mass stars in Carina]{HST/WFC3 Imaging of Protostellar Jets in Carina: [Fe~II] Emission Tracing Massive Jets from Intermediate Mass Protostars}
\author[Megan Reiter and Nathan Smith]{Megan Reiter$^{1}$\thanks{E-mail:
mreiter@as.arizona.edu (MR)} and Nathan Smith$^{1}$ \\
$^{1}$Steward Observatory, Department of Astronomy, University of Arizona, Tucson AZ, 85721, USA}
\begin{document}

\date{Accepted 2013 May 17. Received 2013 May 15; in original form 2012 November 21}

\pagerange{\pageref{firstpage}--\pageref{lastpage}} \pubyear{2013}

\maketitle

\label{firstpage}

\begin{abstract}
We present narrowband WFC3-UVIS and -IR images of four externally irradiated protostellar jets in the Carina nebula: HH~666, HH~901, HH~902, and HH~1066. 
These massive jets are unusual because they are bathed in UV radiation from dozens of nearby O-type stars, but despite the strong incident ionizing radiation, portions of the jet remain neutral. 
Near-IR [Fe~II] images reveal dense, neutral gas that was not seen in previous studies of H$\alpha$ emission. 
We show that near-IR [Fe~II] emitting gas must be self-shielded from Lyman continuum photons, regardless of its excitation mechanism (shocks, FUV radiation, or both). 
High densities are required for the survival of Fe$^+$ amid the strong Lyman continuum luminosity from Tr14, raising estimates of the mass-loss rates by an order of magnitude. 
Higher jet mass-loss rates require higher accretion rates onto their driving protostars, implying that these jets are driven by intermediate-mass ($\sim2-8$ \msun) stars. 
Indeed, the IR driving sources of two of these outflows have luminosities that require intermediate-mass protostars (the other two are so deeply embedded that their luminosity is uncertain). 
All four of these HH jets are highly collimated, with opening angles of only a few degrees, similar to those observed in low-mass protostars. 
We propose that these jets reflect essentially the same outflow phenomenon seen in wide-angle molecular outflows associated with intermediate- and high-mass protostars, but that the collimated atomic jet core is irradiated and rendered observable in the harsh radiative environment of the Carina nebula. 
In more quiescent environments, this atomic core remains invisible, and outflows traced by shock-excited molecules in the outflow cavity give the impression that these outflows have a wider opening angle. 
Thus, the externally irradiated jets in Carina constitute a new view of collimated jets from intermediate-mass protostars, and offer strong additional evidence that stars up to at least $\sim 8$ \msun\ form by the same accretion mechanisms as low-mass stars. 
\end{abstract}


%%%%%%%%%%%%%%%%%% 1. Main Text %%%%%%%%%%%%%%%%%%%%%%%%%%%%%%%%%%%%

\section{Introduction}\label{s:intro}

Herbig-Haro (HH) objects are the emission-line features associated with outflows emanating from young stellar objects (YSOs). 
These optical nebulosities become visible when outflow material collides with the ambient medium or earlier ejecta, thus shock heating the material and leading to the characteristic emission-line spectra \citep{her50,her51,har52,har53}. 
The outflows that produce these nebulosities are formed during a protostar's active accretion phase, and the longest HH flows bear evidence of a protostar's mass-loss history over a significant fraction of its accretion age \citep{rei01}. 

The details of how a jet is launched remain unclear, although energy liberated from mass accretion onto a protostar must power the flow \citep[e.g.][]{cab90,shu94,ouy99,pud06}. 
Jet mass-loss rates from low-mass T Tauri stars appear to be $\sim 1-10$\% of the mass accretion rate \citep{har95,cal98}. 
Thus, the resulting jet morphology and kinematics may provide an indirect record of the protostar's accretion history. 
Outflows are of particular interest for stars at the upper end of the initial mass function (IMF), where the formation mechanism of the highest mass stars remains unclear \citep[e.g.][]{zin07}. 
Jets driven by intermediate-mass stars may be especially illuminating as they sample accretion and outflow in the critical transition between low- and high-mass star formation. 

Most outflows driven by intermediate- to high-mass protostars have been observed via the millimeter emission of entrained molecules or free-free emission from the inner jet \citep[e.g.][]{beu08,for09,fue09,tak12}. 
A wide range of morphologies are seen in the outflows from more massive protostars, from collimated jets \citep[e.g.][]{beu02b} to wide, bubble-like structures, possibly cleared by wide-angle winds \citep[e.g.][]{she99} or even wide-angle explosive phenomena like the Orion BN/KL outflow \citep{bal11}. 
Growing evidence suggests that outflows may be driven by some combination of a collimated jet and a wide-angle wind, with different components dominating at different times \citep[see, e.g.][]{sea08}. 
For more massive protostars (M $\geq$ 2 \msun), \citet{beu05} have proposed an evolutionary scenario where ionizing radiation from the forming star increases the plasma pressure at the base of the outflow to the point that it overwhelms magnetic collimation, leading to wider opening angles. 
Young O-type stars cannot create jet-like outflows in this scenario. 
However, collimated, ionized jets from young, massive protostars have since been observed \citep[e.g.][]{guz11,guz12}, making the relationship between outflow collimation and protostellar mass unclear. 

Young clusters provide an environment where accreting protostars reside alongside massive stars that are already on the main sequence. 
Massive O-type stars bathe those protostars in UV radiation, and this external radiation illuminates the unshocked material in the body of the outflow, allowing for better constraints on the jet properties using diagnostics of photoionized gas. 
UV radiation also erodes the circumstellar environment that would normally obscure an actively accreting protostar.  
Few irradiated jets have been studied in detail, and most that have been studied are driven by low-mass protostars \citep[e.g. in Orion, see][]{rei98,bal00,bal01,bal06}. 
A sample of intermediate-mass protostars caught actively accreting would provide valuable clues to the underlying mechanism that may unify low- and high-mass star formation in the critical $2-8$ M$_{\odot}$ regime. 

In this paper, we study a sample of such objects in the Carina nebula. 
H$\alpha$ observations with \textit{HST} revealed 39 candidate and confirmed HH jets in the Carina nebula with high mass-loss rates, implying that the sample is dominated by intermediate-mass driving sources \citep{smi10a}, although some extremely young, embedded low-mass YSOs are also possible. 
These 39 jets are irradiated by Carina's extreme O-star population, and the photoionization of the jet body allows for an estimate of the jet mass-loss rate from the H$\alpha$ emission measure. 
This is currently the largest known sample of irradiated intermediate-mass flows in a single region, extending the range of masses, environments, and evolutionary stages beyond those seen in Orion. 

Jet properties derived from the H$\alpha$ emission measure provide valuable insights into the nature of the irradiated material in protostellar outflows, but this tracer of the mass still suffers from a number of uncertainties, including the volume filling factor and the ionization fraction in dense regions of the jets. 
While filling factors can be reasonably well estimated from jet morphology in \textit{HST} images, for jets with densities $\geq 2000$ cm$^{-3}$, gas may remain neutral, and so H$\alpha$ may not trace the majority of the mass in the jet, leading to an underestimate of the mass-loss rate \citep[see][]{har94,smi10a}. 

In this paper, we explore the $\lambda 12567$ and $\lambda 16435$ [Fe~II] lines, which probe the low-ionization material in these jets that is not traced by H$\alpha$. 
While photons with $h\nu > 13.6$ eV are absorbed at the ionization front, far-UV radiation ($7.6 - 13.6$ eV) can still penetrate into the dense, neutral portions of these jets, providing a population of singly ionized Fe that traces the otherwise invisible column of neutral atomic gas in the jet. 
In addition, IR wavelengths can penetrate extinction from the globule, in some cases connecting the jet with its optically obscured driving source. 
[Fe~II] $\lambda 12567$ and $\lambda 16435$ both originate from the $a^4D$ level, so their ratio is insensitive to excitation conditions and can be used to estimate variations in the reddening along these jets. 
These [Fe~II] transitions have a high critical density ($n_{crit} \sim 3 \times 10^4$ cm$^{-3}$), and so they trace the dense gas that might make up most of the mass of the jet. 

Here we present a pilot study where we analyse archival WFC3 images of some of the most spectacular HH jets in the Carina nebula --- HH~666, HH~901, HH~902, and HH~1066. 
Narrowband [Fe~II] images provide the first high-resolution IR view of these jets. 
Previous jet mass-loss rates derived from the H$\alpha$ emission measure did not include this dense, neutral gas, and therefore underestimated the mass. 
We discuss these objects as examples of very dense and massive irradiated protostellar jets driven by intermediate-mass protostars.

%%=============================================================================

\section{Observations}\label{s:obs}

\begin{figure*}%[t!]
\centering
$\begin{array}{c}
\includegraphics[angle=0,scale=1.0]{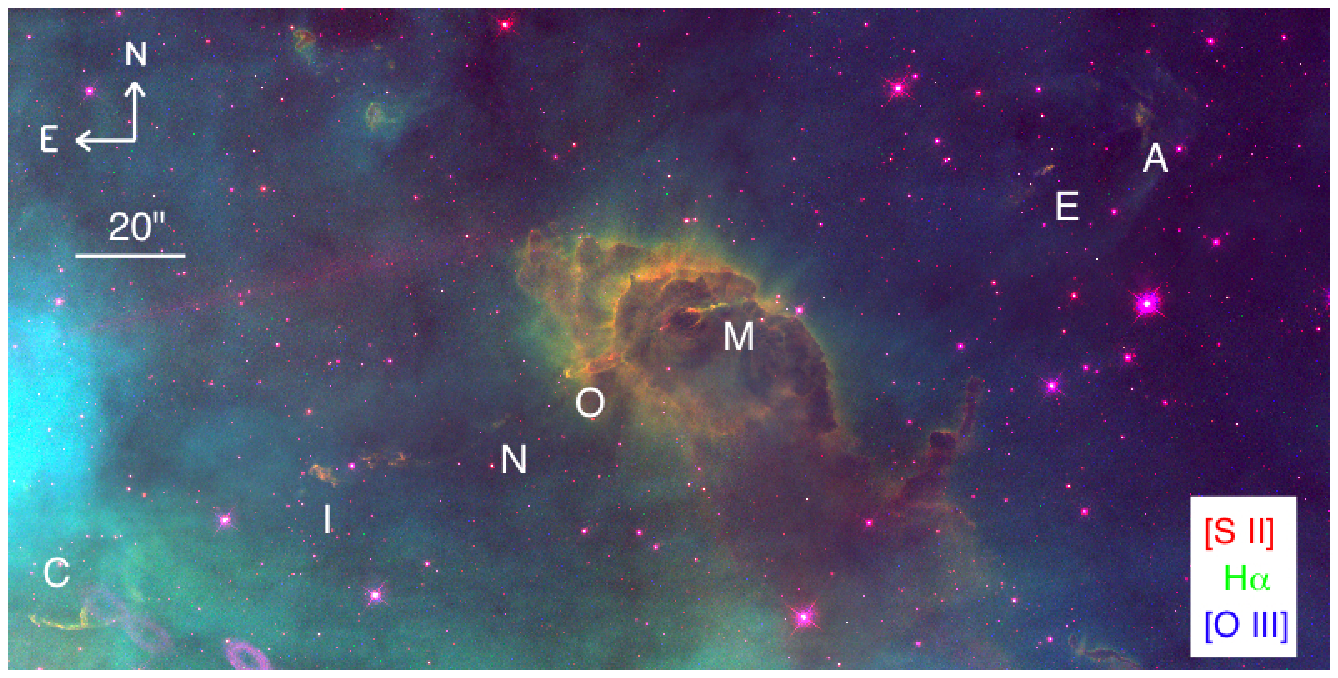} \\
\includegraphics[angle=0,scale=1.0]{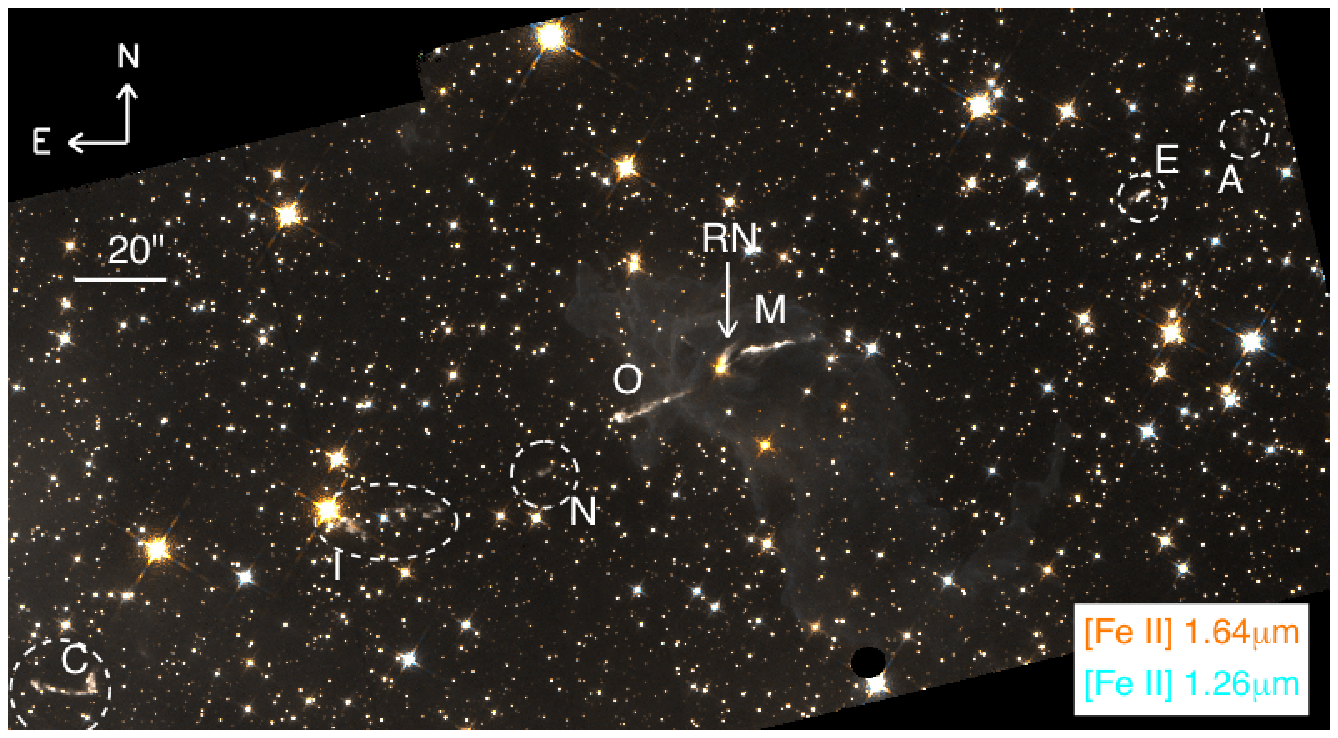} \\
\end{array}$
\caption{WFC3-UVIS (top) and WFC3-IR (bottom) color images of the HH~666 jet. 
Features D, A, E, M, O, N, I, and C identified by \citet{smi04} are the components of the jet; these are labelled here except for feature D which is off the right side of the image. 
Strong emission in both of the observed [Fe~II] lines traces a highly collimated jet back into the parent globule, connecting the optical jet to a bright, embedded intermediate-mass protostar. 
A reflection nebula labelled ``RN'' to the north of the driving source is probably stellar light reflecting off the cavity wall. 
}\label{fig:hh666_full} 
\end{figure*}

After the discovery of many new HH jets in the HST/ACS H$\alpha$ survey of the Carina nebula \citep{smi10a}, a few of the most dramatic jets were targeted with WFC3 for Early Release Observations after SM4, and to commemorate the 20$^{th}$ anniversary of the Hubble Space Telescope. 
WFC3 optical and IR data of HH~666 ($\alpha_{J2000}$ = 10:43:51.54, $\delta_{J2000}$ = -59:55:20.53) were obtained as part of WFC3's Early Release Observations on July 24-30, 2009. 
6\arcmin\ $\times$ 6\arcmin\ mosaics were observed in three narrowband optical filters --- H$\alpha$ (F656N; total integration time 7920s), [S~II] (F673N; 9600s) and [O~III] (F502N; 7920s) --- in addition to [Fe~II] $\lambda 12567$ (600s) and $\lambda 16435$ (700s). No Paschen-$\beta$ images are available for HH~666. 
Angular resolution is a factor of $\sim3$ coarser for the [Fe~II] images compared to the visual wavelength images. 
Drizzled mosaics for HH~666 are available from the Hubble Legacy Archive \footnote{http://hla.stsci.edu/hlaview.html}.

Images of the HH~901 field ($\alpha_{J2000}$ = 10:44:05.25, $\delta_{J2000}$ = -59:29:45.00; the field also includes HH~902 and HH~1066) were taken February 1-2, 2010 to make images for public release in celebration of Hubble's 20th anniversary (PID 12050, P.I. M. Livio). 
Each of the four pointings in the $\sim 6\arcmin \times 6\arcmin$ mosaic was covered with a $2 \times 2$ mosaic pattern with small dithers within each pointing to correct for imaging artifacts. 
Narrowband optical images obtained with WFC3/UVIS complement earlier ACS H$\alpha$ imaging of Trumpler 14 (Tr14), adding a second-epoch H$\alpha$ observation (total integration time 1980s) and high-resolution images in [S~II] (2400s) and [O~III] (3200s).  
In addition to the two [Fe~II] lines (total exposure time of 2400s at $\lambda 12567$ and 2800s at 1.64 \micron), IR images of HH~901 include Pa$\beta$ (F128N; 2800s). 
Unfortunately, no accompanying line-free continuum images in the optical or IR are available from WFC3 for either field (HH~666 or HH~901). 
This complicates some of the analysis because we cannot correct for scattered continuum light that may be included in the narrowband filters. 
As such, we are only sensitive to the brightest [Fe~II] emission. 
In order to extend this analysis to a full statistical sample including fainter HH jets, continuum-subtracted images are required.

%%=============================================================================

\section{Observed Jet Morphology}\label{s:results}

\begin{figure*}%[t!]
\centering
$\begin{array}{cc}
\includegraphics[angle=0,scale=1.0]{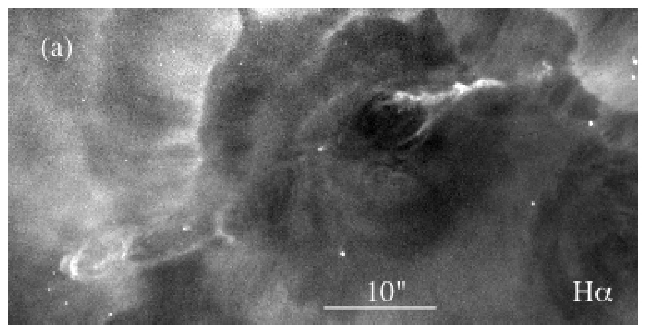} & 
\includegraphics[angle=0,scale=1.0]{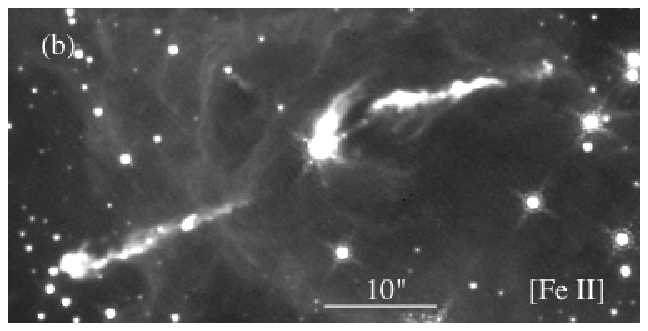} \\ 
\includegraphics[angle=0,scale=1.0]{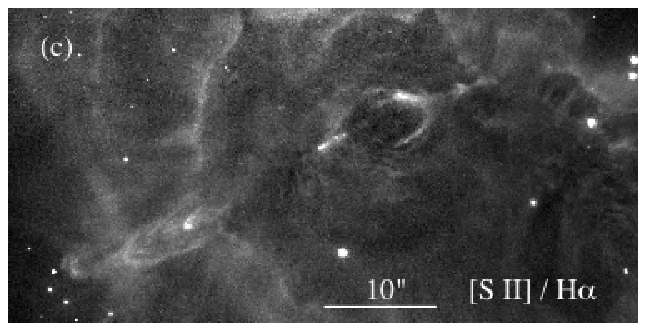} & 
\includegraphics[angle=0,scale=1.0]{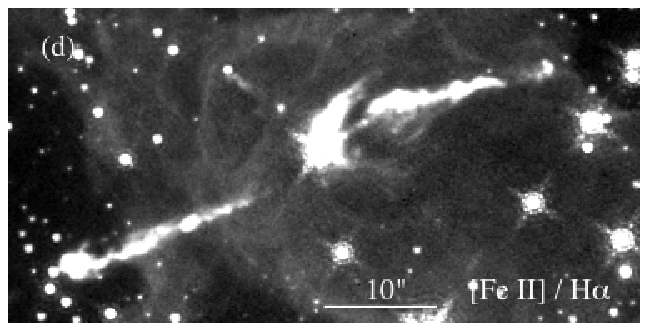} \\
\includegraphics[angle=0,scale=1.0]{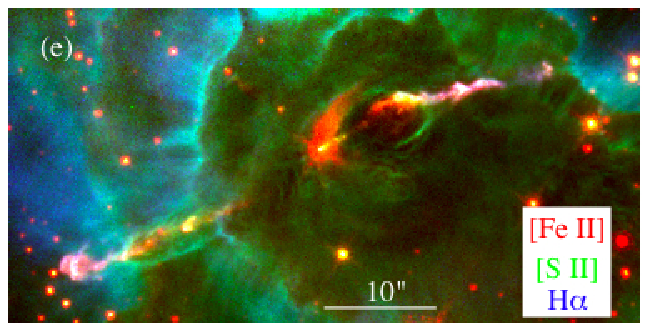} &
\includegraphics[angle=0,scale=0.2575]{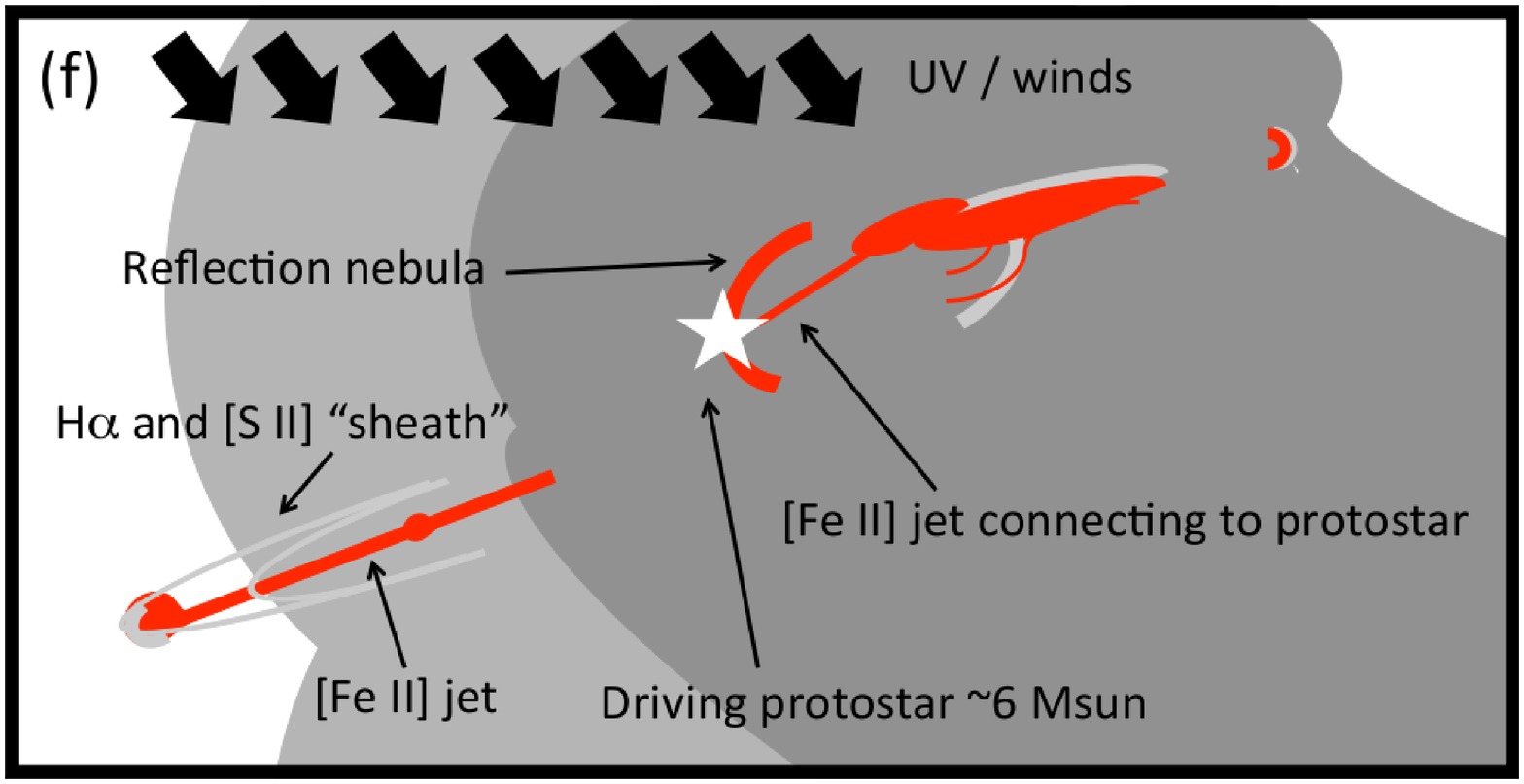}  \\ 
\end{array}$
\caption{Zoomed images of the inner HH~666 jet (features O and M) in H$\alpha$ (a), $\lambda 12657 + \lambda 16435$ [Fe~II] emission (b), [S~II] / H$\alpha$ ratio image (c), $\lambda16435$/H$\alpha$ ratio image (d), a combined color image (e), and a cartoon (f) illustrating the relationship between the optical H$\alpha$ (gray lines) and IR [Fe~II] emission (red lines). 
Highly collimated [Fe~II] emission to the northwest of the IR source clearly connects the jet with the driving protostar. 
H$\alpha$ and [S~II] emission outline [Fe~II] emission in the eastern limb of the jet, possibly tracing material excited in the walls of a cavity cleared by the atomic jet core. 
}\label{fig:hh666_4view} 
\end{figure*}

\subsection{HH~666}\label{ss:hh666}
HH~666 is a spectacular bipolar jet emerging from the head of a dust pillar located $11$ pc south of $\eta$ Carinae \citep{smi04}. 
High resolution H$\alpha$ imaging with HST/ACS confirmed the shock morphology of features HH~666 D, A, E, M, O, N, I, and C \citep[see Figure~\ref{fig:hh666_full}][]{smi10a}. 
Several of the optically identified features in HH~666 also have strong [Fe~II] $\lambda16435$ emission as seen in ground-based images \citep{smi04}. 
The most striking of these features is the clearly delineated bipolar jet embedded within the pillar. 
New images from WFC3 trace the jet back to the driving protostar, and allow us to examine the morphology of the jet in detail. 

The southeast portion of the inner jet (feature HH~666~O) is highly collimated in [Fe~II] emission, with several bright [Fe~II] knots along the flow (Figure~\ref{fig:hh666_4view}b). 
Optical emission traces thin bow shocks that cloak the collimated IR jet in a wider cavity identified as the H$\alpha$ $+$ [S~II] ``sheath'' in Figure~\ref{fig:hh666_4view}. 
The $\lambda16435$/H$\alpha$ flux ratio in the southeast portion of the jet (feature HH~666~O) is over an order of magnitude higher than in the ionization front along the nearby molecular cloud ($\sim 6$ in the jet compared to $\sim 0.2$ in the ionization front).

On the opposite side of the HH~666 driving source, bright optical and IR emission traces a collimated jet (feature HH~666~M, see Figure~\ref{fig:hh666_4view}b).
Bright spots in the [S~II] / H$\alpha$ ratio image (Figure~\ref{fig:hh666_4view}c) trace probable internal working surfaces (bright knots in the middle of HH~666~M and O, see Figure~\ref{fig:hh666_4view}e).  
In contrast, bright emission in the $\lambda 16435/H\alpha$ ratio image (Figure~\ref{fig:hh666_4view}d) is not confined to the two brightest clumps in HH~666~M (where we would expect [Fe~II] to be shock excited), but is bright throughout the full length both HH~666~M and O. 
Thus, bright [Fe~II] emission from the jets does not follow other tracers of shock excitation.

\begin{figure*}
\centering
$\begin{array}{cc}
\includegraphics[angle=0,scale=1.0]{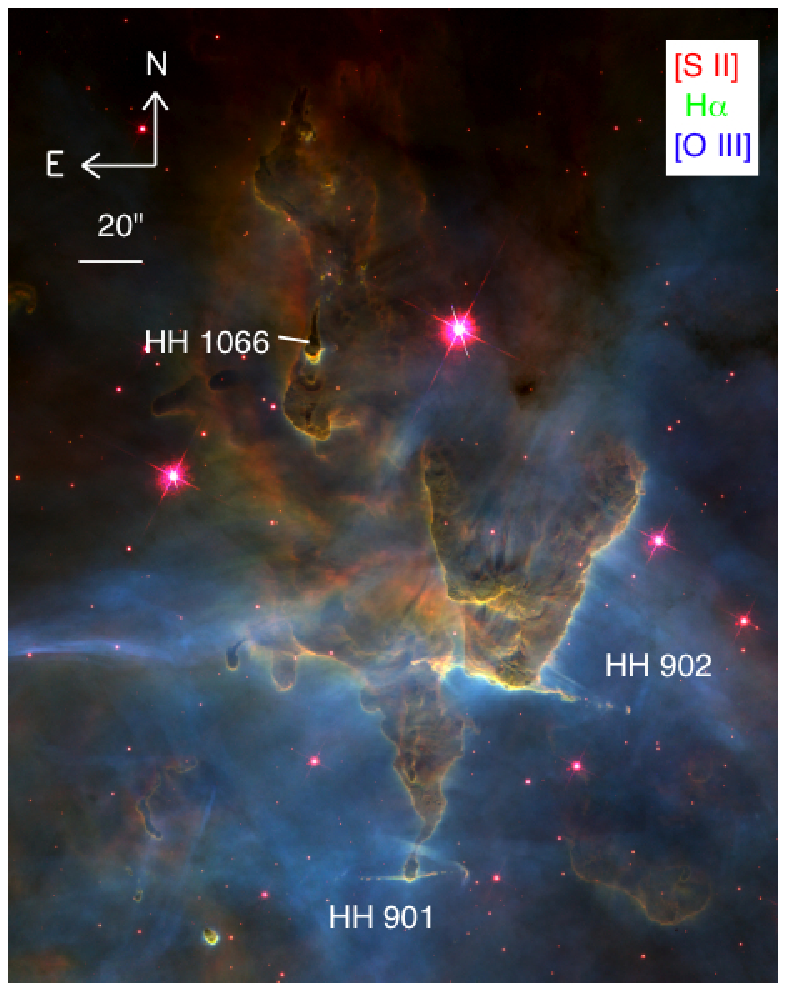} &
\includegraphics[angle=0,scale=1.0]{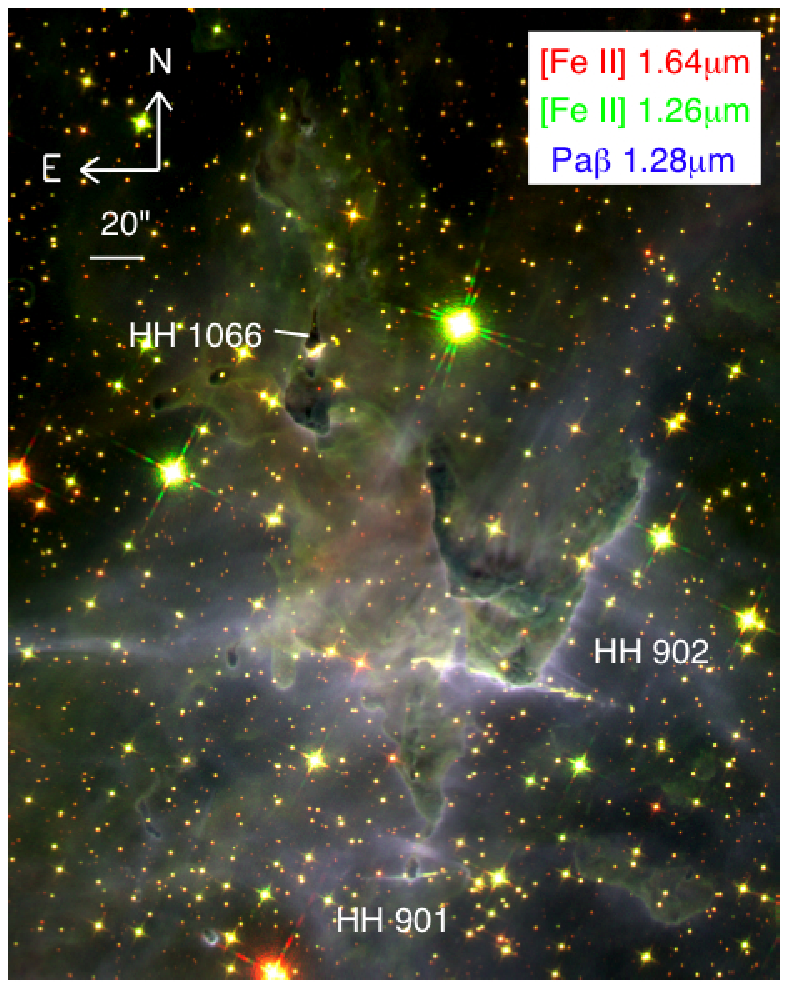} \\
\end{array}$
\caption{Full WFC3-UVIS (left) and WFC3-IR (right) color mosaics of the complex of dust pillars and cometary globules to the northwest of Tr14 housing HH~901, HH~902 and HH~1066. 
[Fe~II] emission allows us to clearly identify the jet morphology of HH~1066 \citep[previously candidate jet HH~c-1,][]{smi10a} and to connect jet emission with a Spitzer-identified intermediate-mass protostar. 
No protostar is identifiable for HH~901 or HH~902, and bright [Fe~II] emission in both jets is only found outside the pillars. 
}\label{fig:hh901_full} 
\end{figure*}

The HH~666 driving source is barely discernible in H$\alpha$ images, but becomes increasingly prominent in the IR. 
Based on analysis of the near-IR SED, \citet{smi04} propose that the driving source (HH~666 IRS) is an embedded intermediate-mass protostar. 
The best-fit model to a more complete IR SED from \citet{pov11} supports this conclusion, reporting a best-fit mass of 6.3 \msun\ (catalog number 345 in their sample).

\subsection{HH~901 and HH~902}\label{ss:hh901_hh902}

HH~901 and HH~902 are neighboring bipolar jets with outflow axes nearly perpendicular to the symmetry axis of their parent clouds \citep[see][]{rag10} and irradiated by the nearby Tr14 star cluster \citep{smi10a}. 
New \textit{HST} images of the field containing HH~901 and HH~902 (as well as HH~1066, formerly candidate jet HH~c-1) are shown in Figure~\ref{fig:hh901_full}. 
Both jets emerge from the heads of dark dust pillars. 
They are highly collimated, with high H$\alpha$ surface brightnesses ($\sim1-3 \times 10^{-14}$ erg/s cm$^{-2}$ arcsec$^{-2}$) that indicate high electron densities of $\gtrsim 1 - 1.7 \times 10^3$ cm$^{-3}$.
Brighter H$\alpha$ emission is found along the south side of the jets facing Tr14, as compared to more diffuse emission from the northern sides of these jets. 
This implies a sharp ionization front along the south-facing edge of the jet body. 
Thus, ionizing photons may not penetrate through the jet, so these jets could have a substantial amount of undetected neutral material. 
They may also be shaped by the stellar winds and/or radiation from Tr14 as discussed in Section~\ref{ss:environment}.

\begin{figure*}%[t!]
\centering
$\begin{array}{c}
\includegraphics[angle=0,scale=1.0]{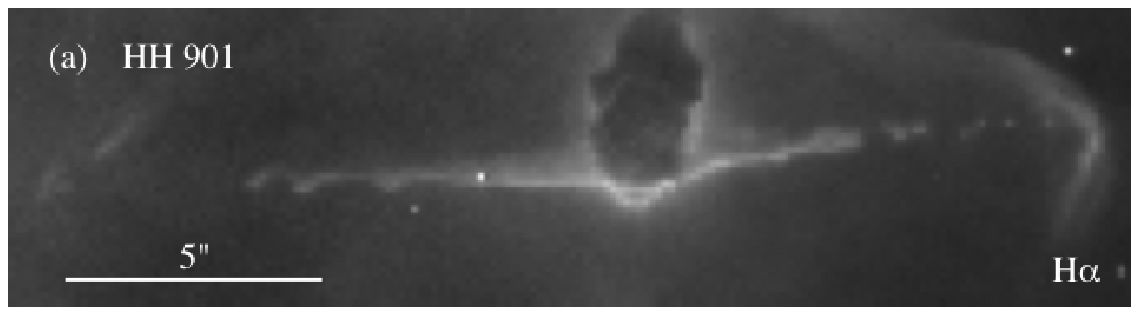} \\
\includegraphics[angle=0,scale=1.0]{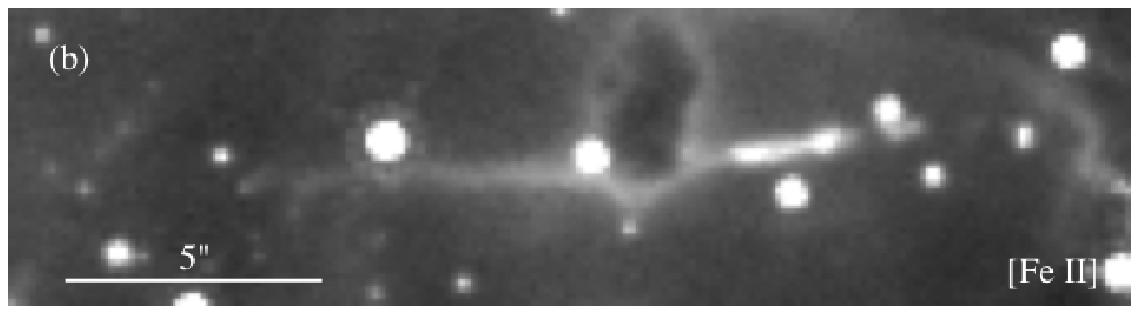} \\ 
\includegraphics[angle=0,scale=1.0]{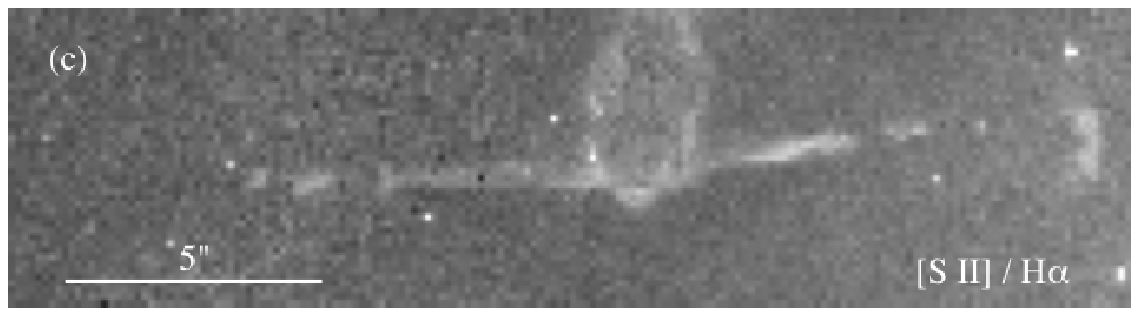} \\ 
\includegraphics[angle=0,scale=1.0]{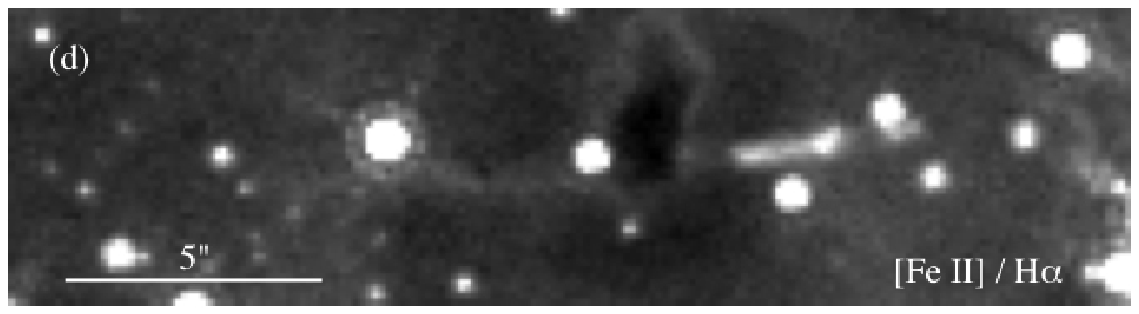} \\ 
\includegraphics[angle=0,scale=1.0]{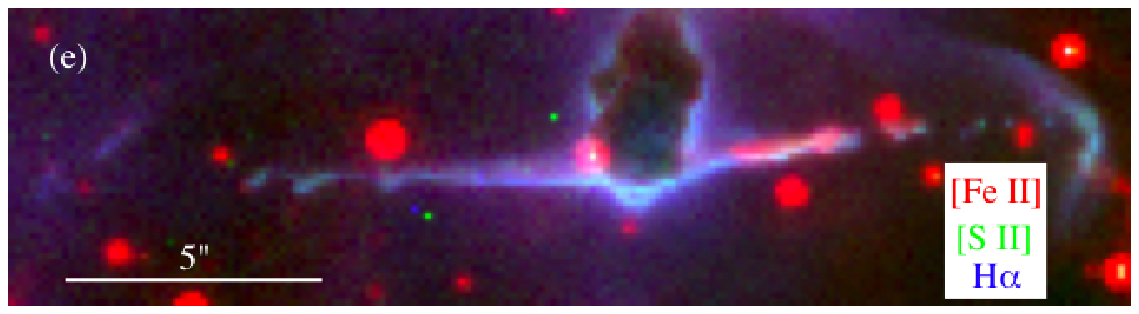} \\ 
\includegraphics[angle=0,scale=0.445]{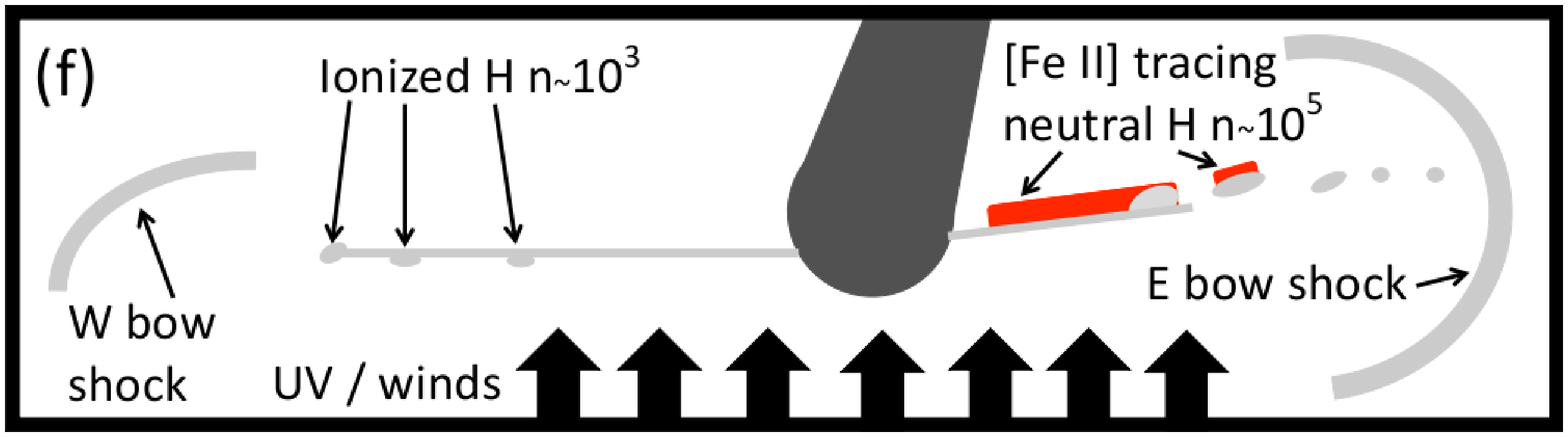} \\ 
\end{array}$
\caption{H$\alpha$ emission (a), $\lambda 12657 + \lambda 16435$ Fe~II] emission (b), [S~II] / H$\alpha$ ratio (c), $\lambda16435$/H$\alpha$ ratio image (d), a color image (e), and cartoon (f) illustrate how [Fe~II] (red lines) peaks \textit{behind} H$\alpha$ (gray lines) in the western limb of the flow. 
}\label{fig:hh901_4view} 
\end{figure*}

\subsubsection{HH~901}\label{sss:hh901}
\begin{figure*}%[t!]
\centering
\includegraphics[angle=0,scale=0.725]{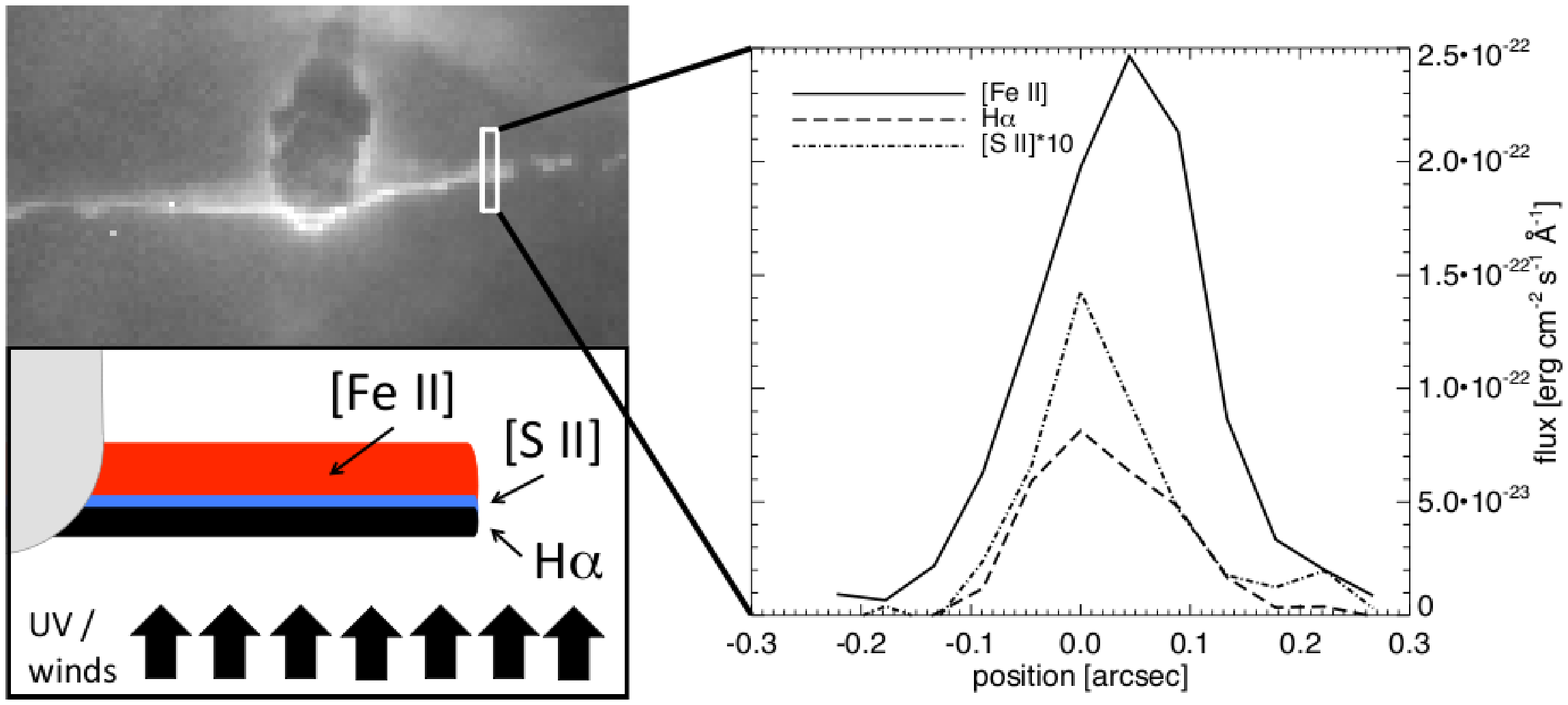}
\caption{\textit{Top Left:} HH~901 H$\alpha$ image indicating location of slice through jet used for the intensity tracing plotted to the right. 
\textit{Bottom Left:} Cartoon of the stratified emission structure of HH~901. Lyman continuum photons ionize H on the side of the jet facing the ionization front (H$\alpha$ emission, black). [S~II] peaks at the same location as H$\alpha$, but is shown slightly offset here (blue). Behind the ionization front, Balmer continuum photons heat the jet, collisionally exciting [Fe~II] emission (red) in the neutral jet core, leading to peak emission behind both optical lines. 
\textit{Right:} H$\alpha$, [S~II], and [Fe~II] $\lambda 16435$ intensity tracings of the western edge of the jet plotted as a function of the offset from the centroid of the H$\alpha$ emission, showing the $0.04\arcsec$ offset between the [Fe~II] and optical line emission. 
}\label{fig:hh901_slice} 
\end{figure*}

Figure~\ref{fig:hh901_4view} provides a detailed view of HH~901 from the new WFC3 images. 
While some [Fe~II] emission is evident throughout the optically identified portions of the jet, [Fe~II] emission is brightest along the western portion of HH~901, offset $\sim 0.5$\arcsec\ from the cometary cloud. 
This portion exhibits a $\lambda16435$/H$\alpha$ flux ratio of $\sim 3$ (compared to $\sim 7 \times 10^{-4}$ from the ionization front along the neighboring pillar). 
Extended [Fe~II] emission stretches along the jet axis, but its centroid is offset $\sim 0.05$\arcsec\ north of the thin H$\alpha$ and [S~II] emission at the edge of the jet (see Figure~\ref{fig:hh901_slice}). 

A sketch of the observed stratified emission structure in HH~901 is presented in Figure~\ref{fig:hh901_slice}. 
For a sufficiently dense jet in an external radiation field, an ionization front will exist in the jet itself, resulting in an ionized skin shielding a core of neutral H (Figure~\ref{fig:hh901_4view}e,f). 
[S~II] emission in the jet does not extend north of the narrow ionization front to where [Fe~II] emission is brightest, probably resulting from high densities ($>10^3$ cm$^{-3}$) in the neutral jet where [S~II] is collisionally de-excited. 

Unlike in HH~666, [Fe~II] emission does not trace HH~901 all the way back to the driving source. 
In fact, [Fe~II] emission from HH~901 is only detectable outside the pillar boundary. 
Possible explanations for this difference include optical depths in the HH~901 pillar so high that external FUV photons cannot penetrate the cloud (so that [Fe~II] emission is not excited), extinction blocks the inner jet, or that the driving protostar is no longer actively accreting and therefore no longer driving a jet. 
A steep drop in the accretion rate would need to be timed correctly so that a drop in density along the jet exactly coincides with the edge of the cloud. 
Assuming $V= 200$ km s$^{-1}$, this would have had to happen within the last $\sim 55$ years, making this possibility unlikely. 
We discuss the other two potential explanations in turn below.  

The well-defined ionization front along the edge of the HH~901 pillar suggests very high densities ($\geq 3 \times 10^4$ cm$^{-3}$, see Section~\ref{ss:mdot_nd}) such that Lyman continuum radiation cannot penetrate beyond the pillar edge. 
Conversely, the ionization front along the HH~666 pillar is significantly less sharp with evaporative streams extending $\sim 5- 10\arcsec$ from the ionization front. 
The edges of both pillars are bright in the [S~II] / H$\alpha$ ratio image, indicating a significant drop in the ionization fraction at the edge of the pillar \citep[Figure~\ref{fig:hh901_4view}c,][]{dom94}. 
The central protostar is not seen in HH~901, but is clearly seen in HH~666. 
Moreover, several background stars shine through the pillar hosting HH~666 in the near-IR images. 
Together, this suggests that the inner regions of HH~901 are obscured by dust in a globule that is more dense and opaque than the comparatively ``fluffy'' HH~666 pillar. 
Flux limits indicate that any [Fe~II] emission from within the HH~901 pillar must be at least $\sim 400$ times fainter than in the western limb of the jet. 
If the inner jet has the same brightness as in the western limb, this implies $\geq 6$ mag of extinction at $\lambda 12567$ ($A_V \approx 9$ mag), or a column of $ \gtrsim 2 \times 10^{22}$ cm$^{-2}$ \citep{guv09}.
This is in contrast to the HH~666 pillar, where the $\lambda 16435/H\alpha$ ratio suggests $\tau \approx 0.2$ at 1.26 \micron, corresponding to $A_V \approx 0.3$ mag (see Figure~\ref{fig:feii_ebv}, Section~\ref{ss:hh1066}). 
With the smaller physical size of the pillar housing HH~901 ($\sim 2$\arcsec\ wide, $\sim0.02$ pc at a distance of $2.3$ kpc) than the HH~666 pillar ($\sim 40$\arcsec\ wide, $\sim 0.4$ pc), this implies much higher volume densities obscuring the HH~901 driving source ($n_H \approx 3 \times 10^{5}$ cm$^{-3}$ compared to $n_H \approx 4 \times 10^{2}$ cm$^{-3}$ for HH~666).

\begin{figure*}%[h!]
\centering
$\begin{array}{c}
\includegraphics[angle=0,scale=1.0]{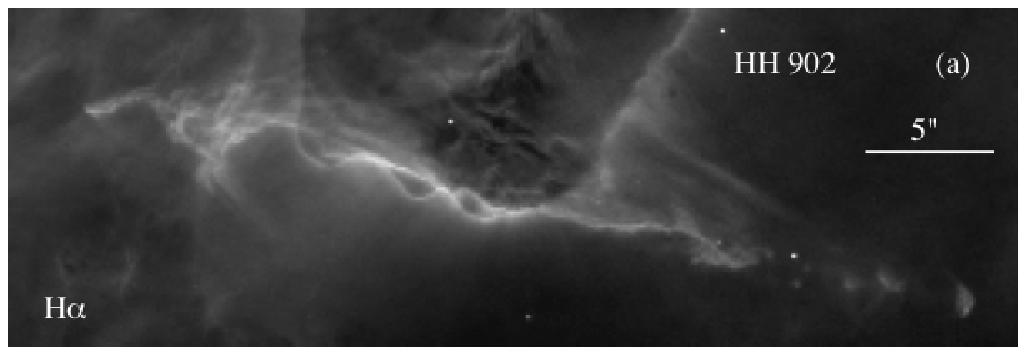} \\
\includegraphics[angle=0,scale=1.0]{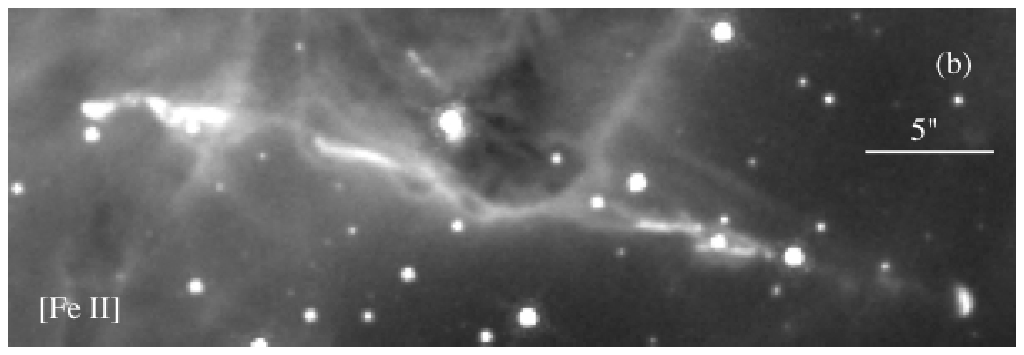} \\ 
\includegraphics[angle=0,scale=1.0]{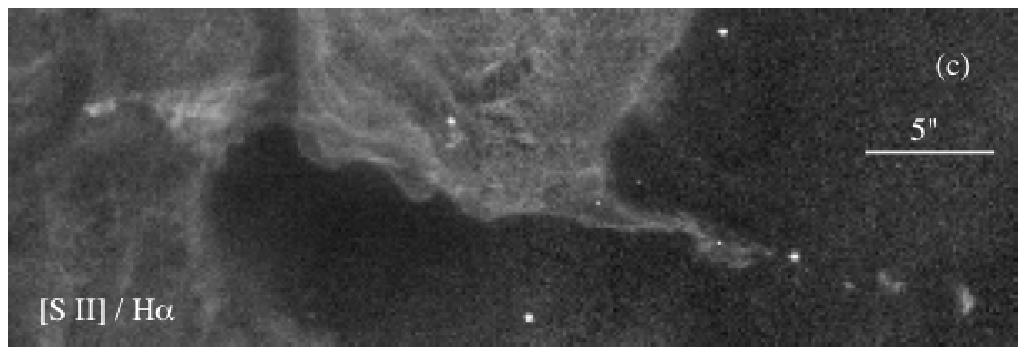} \\ 
\includegraphics[angle=0,scale=1.0]{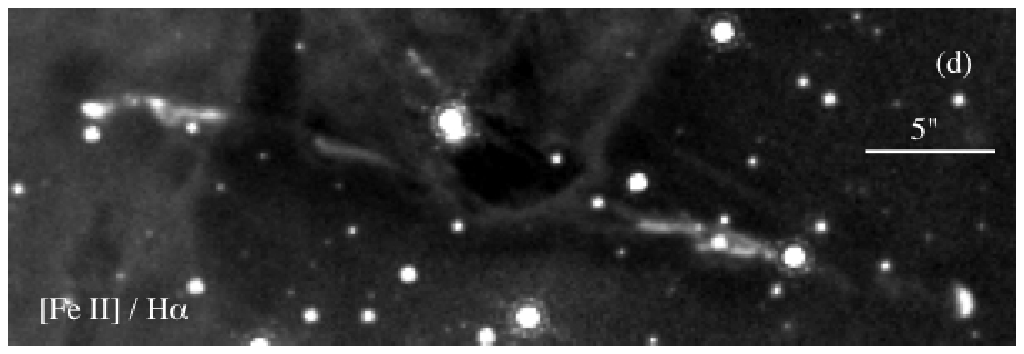} \\ 
\includegraphics[angle=0,scale=1.0]{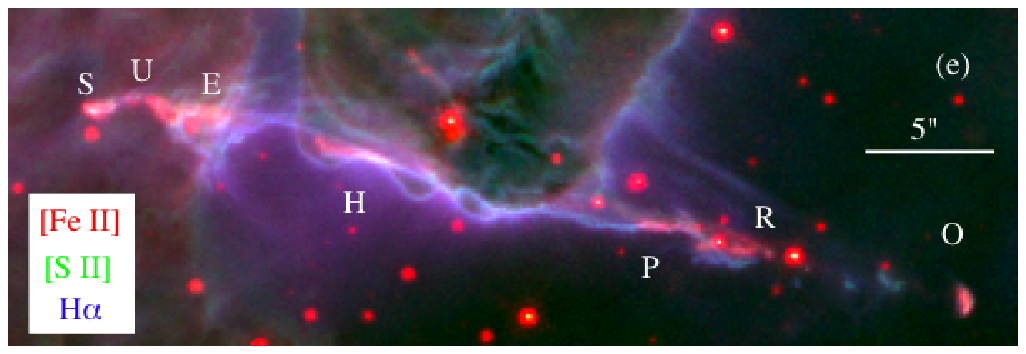} \\ 
\includegraphics[angle=0,scale=0.625]{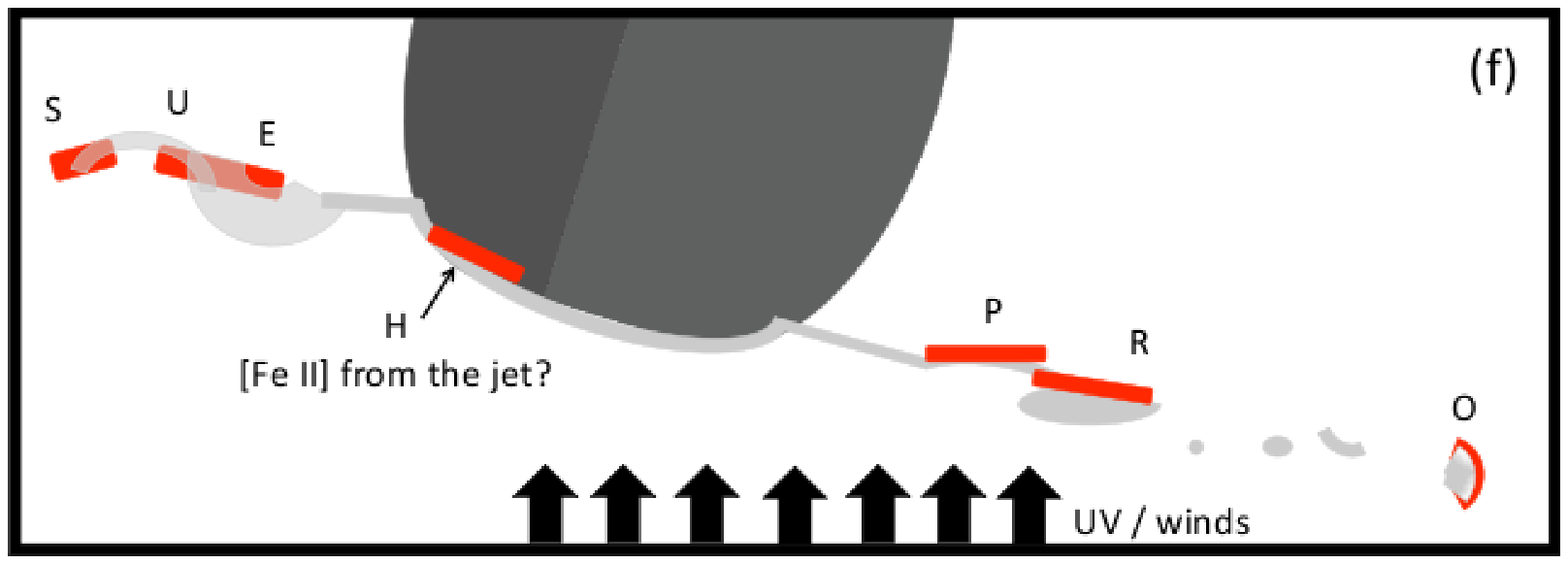} \\ 
\end{array}$
\caption{Detailed view of HH~902 with H$\alpha$ (a), $\lambda 12657 + \lambda 16435$ [Fe~II] emission (b), [S~II] / H$\alpha$ ratio (c), $\lambda16435$/H$\alpha$ ratio (d), a color image (e), and cartoon (f) illustrate the stratified emission structure of the jet (red lines illustrate [Fe~II] emission while gray lines show H$\alpha$). 
Bright [Fe~II] emission features and remarkable changes in the optical jet are labeled as O, R, P, H, E, U, and S and are discussed in Section~\ref{sss:hh902}.  }\label{fig:hh902_4view} 
\end{figure*}

No point-like emission is detected from the tip of the HH~901 pillar in the WFC3-IR bands, nor at longer wavelengths with Spitzer and Herschel \citep{pov11,ohl12}. 
Bright 24 \micron\ emission from the surroundings makes it impossible to confirm or rule out any point-like emission that would signify the presence of an extremely young, deeply embedded low- or intermediate-mass protostar. 
More detailed analysis is needed and will be the subject of a follow-up paper \citep{reiprep}. 
The lack of emission at all observed IR wavelengths supports the conjecture that we do not observe [Fe~II] emission from the jet within the pillar due to high dust optical depths at $1-2$ \micron. 

\begin{figure*}%[t!]
\centering
$\begin{array}{ccc}
\includegraphics[angle=0,scale=1.0]{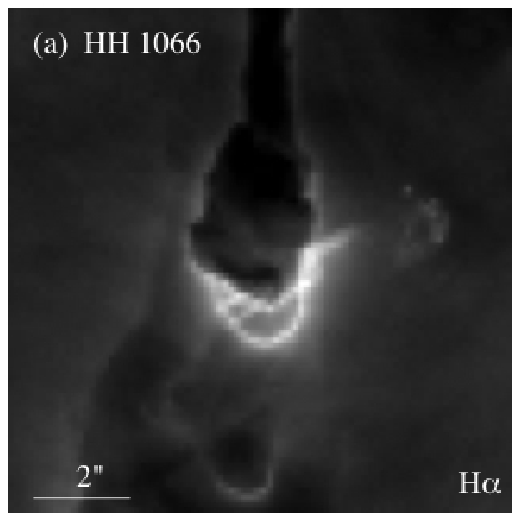} & 
\includegraphics[angle=0,scale=1.0]{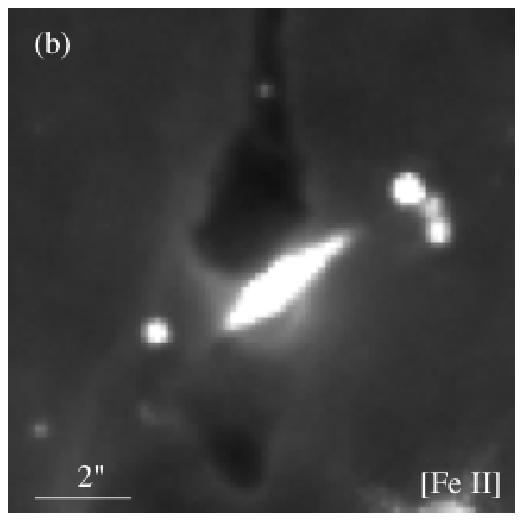} & 
\includegraphics[angle=0,scale=1.0]{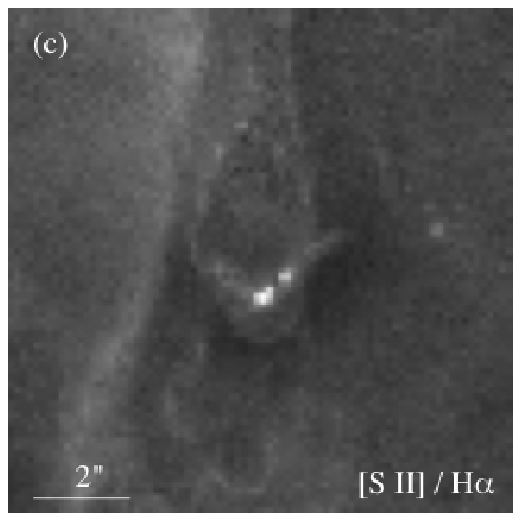} \\
\includegraphics[angle=0,scale=1.0]{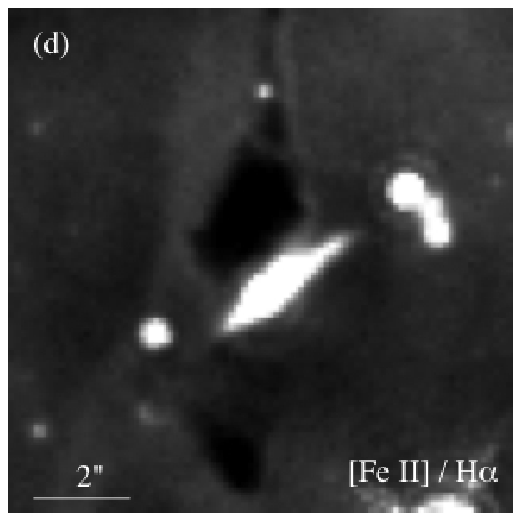} &
\includegraphics[angle=0,scale=1.0]{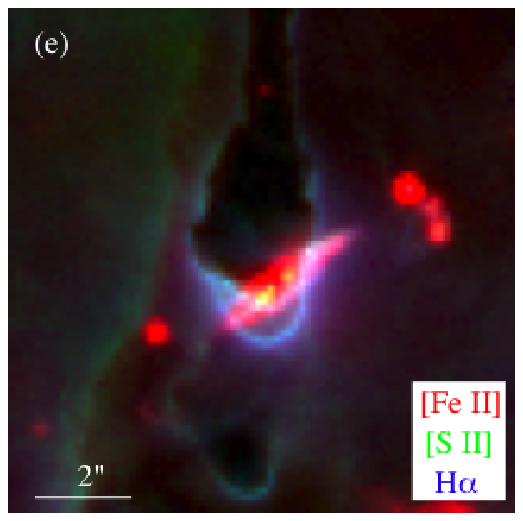} &
\includegraphics[angle=0,scale=0.275]{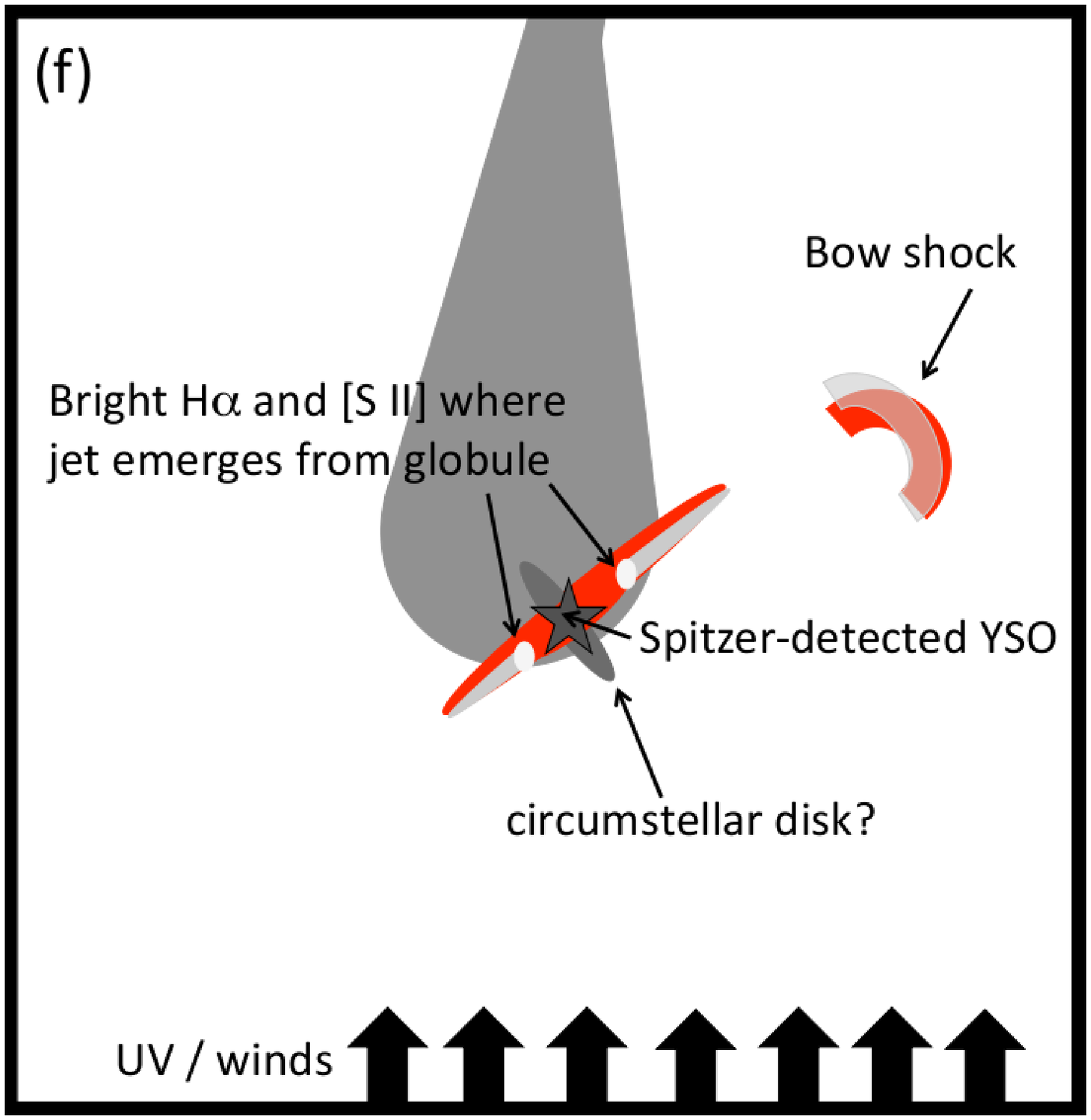}  \\ 
\end{array}$
\caption{Detailed view of HH~1066, which emerges from a small globule near HH~901 and HH~902. 
H$\alpha$ (a) and [Fe~II] $\lambda 12657 + \lambda 16435$ (b) clearly reveal the jet morphology. 
Two bright spots in the [S~II] / H$\alpha$ ratio image (c) reveal where the jet punches out of the globule. 
The $\lambda 16435$/H$\alpha$ ratio image (d), and a combined color image (e). 
A cartoon (f) illustrates our interpretation that optical emission (H$\alpha$, gray lines) near the driving source is obscured by an optically thick disk ([Fe~II] emission is shown in red). Optical and IR emission from the jet are shown slightly offset for clarity.}
\label{fig:hh1066_4view} 
\end{figure*}

\subsubsection{HH~902}\label{sss:hh902} 
The clumpy structure of HH~902 is evident in multiple knots of bright [Fe~II] emission from the jet (Figure~\ref{fig:hh902_4view}), which we label as HH~902 O, R, P, H, E, U, and S. 
Emission from the western limb of the flow is bright beyond the pillar edge ($\sim 1$\arcsec), revealing two narrow features along the jet axis and bright, arc-like [Fe~II] emission from the bow shock (feature O, Figure~\ref{fig:hh902_4view}). 
As in HH~901, the centroid of narrow [Fe~II] is offset from H$\alpha$ and grows bright $\sim 0.2\arcsec$ north of the ionization front traced by H$\alpha$ and [S~II] (feature R). 

Bright [Fe~II] along the eastern edge of the pillar (feature H) is probably the jet skimming the pillar edge. 
The high $\lambda 16435/H\alpha$ ratio here 
%($\sim 1.75$) 
indicates that [Fe~II] emission comes from the jet emerging from the pillar, rather than the ionization front and cloud edge ($\lambda 16435/H\alpha \sim 2$ from the jet compared to $\lambda 16435/H\alpha \sim 0.5$ along the adjacent pillar edge). 

To the east, [Fe~II] emission from the jet is relatively narrow ($\sim 0.5\arcsec$). 
Features E, U, and S have $\lambda 16435/H\alpha$ $\sim 5$, $\sim 6$, and $\sim 8$ respectively, much higher than in any nearby cloud. 
As with HH~901, there is no clear evidence for [Fe~II] emission from the HH~902 jet base or near-IR continuum from the driving source, suggesting that the HH~902 pillar is similarly dense and opaque.

\subsection{HH~1066}\label{ss:hh1066}

\citet{smi10a} listed a possible collimated jet emerging from a cometary cloud near HH~901 and HH~902 as candidate jet HH~c-1. 
New WFC3 optical and IR images (Figure~\ref{fig:hh1066_4view}) clearly show a collimated bipolar flow emerging from the apex of the cloud. 
[Fe~II] emission is too bright to be consistent with structure in the ionization front, with $\lambda 16435$/H$\alpha > 30$. 
We therefore assign an HH number of 1066 to this protostellar jet. 
The newly-minted HH~1066 is striking in the [S~II] and H$\alpha$ images, as they reveal two bright spots where the jet punches out of the parent cloud. 
These two spots lay close together ($\sim 0.4\arcsec$) and have roughly equal brightness, suggesting a nearly edge-on orientation for the disk-jet system, with an optically thick circumstellar disk obscuring the base of the jet in all three of the optical bands. 

[Fe~II] emission at both $\lambda 12657$ and $\lambda 16435$ is extremely bright throughout the HH~1066 jet body and in the bow shock on the western edge of the flow (see Figure~\ref{fig:hh1066_4view}). 
Some fainter [Fe~II] emission opposite this feature to the east may be the complementary bow shock, but proper motions are required to confirm this. 

Both [Fe~II] $\lambda12567$ and $\lambda16435$ originate from the same upper level (a$^4$D), so their intrinsic flux ratio is determined by atomic physics \citep[see, e.g.][]{sh06}. 
Observed deviations from that ratio provide a direct measure of the reddening. 
Ideally, a flux ratio would be done with continuum-subtracted images, but unfortunately, no narrowband WFC3 continuum images are available for this epoch because these images were obtained to make public outreach images.
However, HH~1066 is in a less complicated environment than the other three jets, making it possible to fit and subtract the sky emission around the jet from the [Fe~II] images themselves. 
Local sky emission for each position along the jet is estimated with linear fits to sky emission adjacent to the jet. 
After subtracting estimated sky emission, we measure the ratio $\mathcal{R} = \lambda16435$/$\lambda12567$. 
In the absence of any extinction, the $\lambda 16435$ line will be fainter \citep[][derive an intrinsic $\mathcal{R} = 0.67$]{sh06}. 
An increase in $\mathcal{R}$ will correspond to an increase in the reddening. 

\begin{figure}%[t!]
\centering
$\begin{array}{c}
\includegraphics[angle=0,scale=1.0]{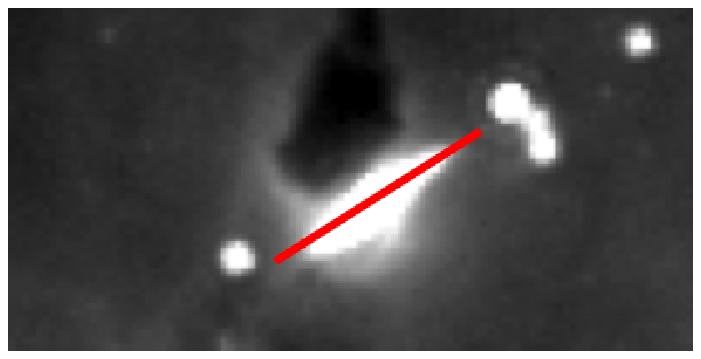} \\
\includegraphics[angle=90,scale=0.325]{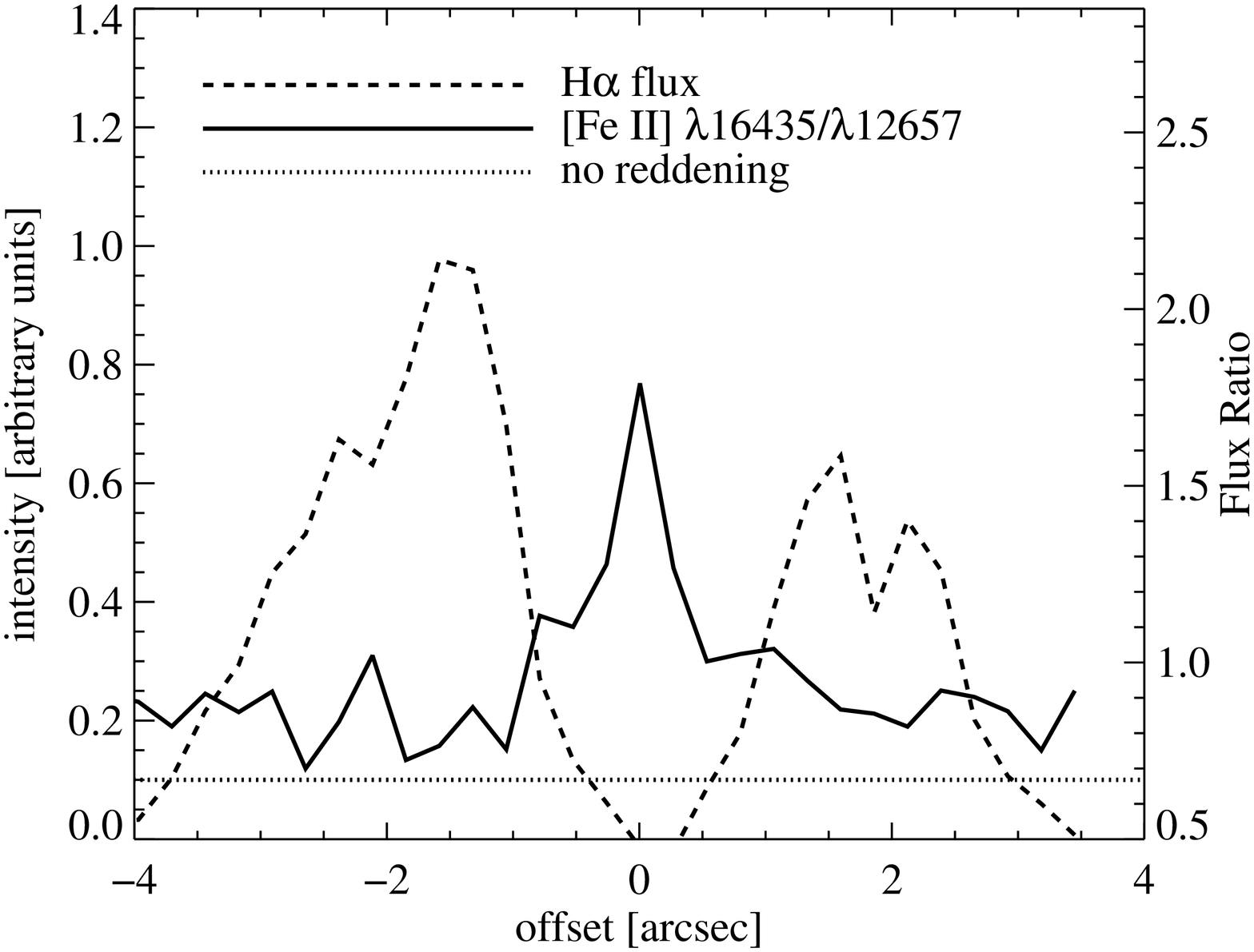} \\
\end{array}$
\caption{\textit{Top:} An F164N WFC3-IR image of HH~1066 showing the location of the jet cross-section used to make the [Fe~II] ratio plot shown in the bottom panel. 
\textit{Bottom:} Intensity tracings of H$\alpha$ and the ratio of [Fe~II] $\lambda 16435$ / $\lambda 12567$ across the length of HH~1066. 
The intensity of the [Fe~II] ratio reaches a maximum where H$\alpha$ intensity is a minimum, suggesting an increase in the optical depth. 
}\label{fig:hh1066_feIIrat} 
\end{figure}

\begin{figure}%[t!]
\centering
\includegraphics[angle=90,scale=0.325]{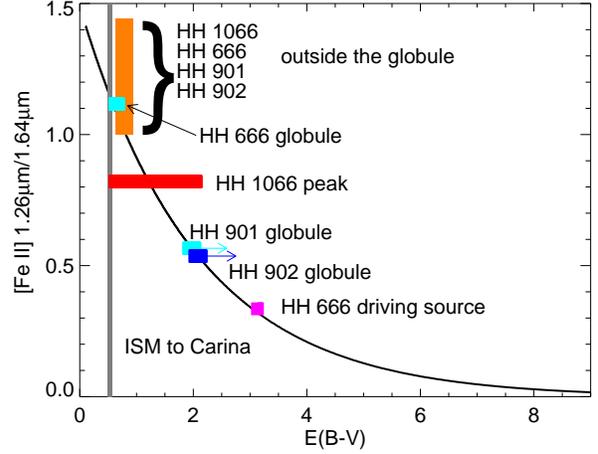} 
\caption{Measured [Fe~II] 1.26 \micron\ / $\lambda 16435$ ratio \citep[intrinsic value is 1.49, see][]{sh06} as a function of E(B-V) for $R=4.8$ \citep[as measured toward Carina;][]{smi87,smi02}. 
The peak of the [Fe~II] ratio in Figure~\ref{fig:hh1066_feIIrat} corresponds to E(B-V) $\approx 2.5$ mag (note that the ratio in Figure~\ref{fig:hh1066_feIIrat} is the inverse of that plotted here). 
}\label{fig:feii_ebv} 
\end{figure}

Figure~\ref{fig:hh1066_feIIrat} shows $\mathcal{R}$ measured along the HH~1066 jet. 
We find $\mathcal{R}$ is relatively constant throughout most of the jet, but the ratio increases toward the jet base (within $2\arcsec$), reaching a maximum of $\mathcal{R} = 1.22$. 
Using $R=4.8$, as measured toward Carina \citep{smi87,smi02}, this is equivalent to an $A_V \approx 13$ mag, where optical emission from the jet disappears (see Figures~\ref{fig:hh1066_feIIrat} and~\ref{fig:feii_ebv}). 
This is morphologically consistent with an edge-on flared circumstellar disk blocking optical radiation around the jet base. 
$A_V$ may increase further in narrow regions along the disk where [Fe~II] is obscured as well\footnote{At 1.64 \micron, the angular resolution is 0.165\arcsec, corresponding to $\sim 400$ AU, over which the optical thickness of the disk may be expected to change significantly \citep[see, e.g.][]{cun11}.}. 

Model fits to the SED of a Spitzer-detected YSO located along the HH~1066 jet axis yield a stellar mass estimate of 2.8 \msun\ \citep[catalog number 429 from][]{pov11}. 
The jet mass-loss rate of $\sim 10^{-7}$ \msun\ yr$^{-1}$ estimated from the H$\alpha$ emission measure \citep{smi10a} implies an accretion rate $> 10^{-6}$ M$_{\odot}$ yr$^{-1}$ (assuming $\dot{M}_{jet} / \dot{M}_{acc} = 0.1$), two orders of magnitude higher than average accretion rate found for intermediate-mass T Tauri stars \citep[$\sim 3 \times 10^{-8}$ M$_{\odot}$ yr$^{-1}$,][]{cal04}, suggesting that this source is very young. 

%%=============================================================================

\section{Discussion}\label{s:discussion}

\subsection{[Fe~II] Emission as a Tracer of Dense, Self-Shielded Gas}\label{ss:about_feII}
We detect extended emission in the F126N and F164N filters that indicates strong [Fe~II] $\lambda 12567$ and $\lambda 16435$ emission in all four of the HH jets in the Carina nebula imaged with WFC3-IR. 
In HH~666 and HH~1066, [Fe~II] emission traces the jet back to the embedded driving source. 
For both HH~901 and HH~902 however, [Fe~II] emission in the jet is only detected outside the boundaries of the parent pillar. 

Given the strong ionizing photon flux of the $\sim 70$ O-type stars in the Carina nebula, it is surprising to find jets inside the H~{\sc ii} region that retain a substantial fraction of neutral gas. 
The gas in the jet must be very dense in order to self-shield, and we find that the IR [Fe~II] emission is an excellent tracer of this self-shielded gas. 

Emission in [Fe~II] $\lambda12567$ and $\lambda16435$ is often assumed to be shock excited since it is seen in supernova remnants \citep[e.g.][]{mor02}, but shock excitation does not necessarily dominate in regions with significant FUV radiation. 
When infrared [Fe~II] emission is produced by photoexcitation, it requires strong FUV radiation, but the gas must be shielded from Lyman continuum in order to prevent ionization to Fe$^{++}$ (the Fe$^+$ ionization potential is $16.2$ eV). 
Regardless of the excitation mechanism for the [Fe~II] emission (shocks or FUV radiation), the fact that Fe$^+$ survives provides a key constraint on the density in the jets (see Section~\ref{ss:mdot_nd}). 
Furthermore, the [Fe~II] surface brightness in these jets is an order of magnitude or more higher than that measured in the ionization fronts in the dusty pillars from which the jets emerge, demonstrating that they are not simply extensions of the pillar ionization front. 

\citet{rei00} speculate that FUV emission from the driving protostar is the most important source of FUV radiation for their sample of seven HH jets driven by low mass stars that are located in quiescent environments. 
If FUV from the protostars were the main source of ionizing radiation for these four jets in the Carina nebula, then we would expect [Fe~II] to be brightest near the driving sources. 
While this may be the case for HH~1066, in the other three jets the brightest [Fe~II] emission is offset from the driving source ($\sim 4$\arcsec\ for HH~666) or from clumps beyond the edge of the pillar (HH~901 and HH~902). 
The clumps nearest to the jet origin do not appear substantially brighter than those further out. 

Observations of other non-irradiated HH jets generally find [Fe~II] emission confined to knots in the outflow with morphology similar to other shock tracers \citep[e.g. H$_2$ 2.12 \micron, and He~{\sc i} 1.083 \micron,][respectively]{lor02,tak02}. 
The excitation energy and critical density of [Fe~II] is similar to (but slightly higher than) that of [S II] and emission from the two shock-tracers is often coincident \citep[see, e.g.][]{tak02,pod06}. 
However, this is not the case in the HH jets in Carina where [S~II] emission is nearly identical to the H$\alpha$ morphology and does not appear to be especially bright in any of the [Fe~II] knots. 
Clumpy but elongated [Fe~II] $\lambda 16435$ emission has been seen in other externally irradiated HH jets \citep[e.g.][]{ell13}. 

HH~901 and HH~902 display a stratified ionization structure traced by H$\alpha$, [S~II], and [Fe~II] emission, with a decreasing ionization fraction with increasing distance from Tr14 (Figure ~\ref{fig:hh901_slice}). 
This suggests that these jets are heated externally by UV radiation from Tr14. 
H$\alpha$ emission traces the ionization front that protects a dense, largely neutral portion of the jet revealed only by a high density, low ionization tracer like [Fe~II]. 
In HH~901, the [S~II] / H$\alpha$ ratio \citep[which increases with decreasing ionization fraction, see][]{dom94} is highest along the western limb where [Fe~II] is also brightest, suggesting a sharp boundary in the ionization front (see Figure~\ref{fig:hh901_4view}).

Bright [Fe~II] emission resides behind the ionization front, and traces the densest neutral material that does not emit H$\alpha$, requiring that these jets are more massive than previously thought. 
Existing mass-loss rate estimates use the electron densities derived from the H$\alpha$ emission measure and assume that the HH jets in the Carina nebula are fully ionized \citep{smi10a}. 
However, H$\alpha$ emission only traces ionized gas, not the total gas mass. 
In the following section, we reevaluate the jet density and mass-loss rate, taking into account this previously unseen neutral material. 

\subsection{Jet Mass and Density Estimates}\label{ss:mdot_nd}
The detection of bright [Fe~II] emission that is systematically offset from H$\alpha$ emission in a direction away from the source of ionizing photons (Tr14) suggests that a significant fraction of the jet mass is shielded and exists as neutral hydrogen, as hypothesized by \citet{smi10a}. 
The fact that this gas is not fully ionized, despite the harsh UV environment, means that it must be very dense. 
To quantify this, we derive jet density and mass-loss rates for the eastern and western limbs of HH~901 without assuming that the jet is fully ionized.  

\citet{bal06} define a minimum density for the jet to be optically thick to Lyman continuum radiation, such that an ionization front is created at a surface within the jet. 
To maintain a slow-moving ionization front (rather than rapidly ionizing the entire jet body), the average atomic density in the neutral jet beam, $n_H$, must be greater than a minimum density, 
\begin{equation}
n_{min} = \left| \frac{F}{2 \alpha_B r(d)} \right| ^{1/2} = \sqrt{ \frac{Q_H}{4 \pi D^2} \frac{\mathrm{sin}(\beta)}{2 \alpha_B r} }
\end{equation} 
where $F$ is the ionizing flux from Tr14, which has a total ionizing photon luminosity log($Q_H$) $= 50.34$ photons s$^{-1}$ \citep{smi06}.  
For HH~901, $D = 0.7$ pc is the projected distance from Tr14 to the jet (the distance may be somewhat larger due to uncertainty in the 3D geometry), 
$\beta$ is the angle between the jet axis and a line to the illuminating source (taken to be 90$^{\circ}$ here; smaller angles between HH~901 and Tr14 will decrease $n_{min}$), 
$\alpha_B \approx 2.6 \times 10^{-13}$ cm$^3$ s$^{-1}$ is the case B recombination coefficient for hydrogen, and 
$r$ is the radius of the cylindrical jet column. 
For HH~901, $n_{min} \approx 3.4 \times 10^4$ cm$^{-3}$, which is very similar to the critical density of these [Fe~II] transitions ($n_{crit} \sim 3 \times 10^4$ cm$^{-3}$), but larger than the electron density derived from the H$\alpha$ emission measure ($n_e = 0.9 - 1.4 \times 10^3$ cm$^{-3}$). 

From $n_{min}$, we can calculate a lower limit on the mass-loss rate in a cylindrical jet using 
\begin{equation}
\dot{M} = n_{min} | \pi \mu m_H V r^2 |
\end{equation}
where 
$\mu$ is the mean molecular weight ($\approx 1.35$), 
$m_H$ is the mass of hydrogen, and
$V$ is the jet velocity. 
This can be expressed numerically as 
\begin{equation}
\dot{M}_{min} = 6.2 \times 10^{-10} \left( \frac{n_{min}}{3.4 \times 10^4 cm^{-3}} \right) \, V_{200} \, r_{pc}^{2} \, M_{\odot} \, yr^{-1}
\end{equation}
where 
$n_{min}$ is the density in units of $3.4 \times 10^4$ cm$^{-3}$, 
$V_{200}$ is the jet velocity in units of 200 km s$^{-1}$, 
and $r_{pc}$ is expressed in pc. 
For HH~901, the minimum mass-loss rate for the jet to be optically thick to Lyman continuum photons is $\dot{M}_{min} = 2.3 \times 10^{-6}$ M$_{\odot}$ yr$^{-1}$. 

An independent way to estimate $\dot{M}$ is to compare the jet mass-loss rate to the photoablation rate, such that a neutral cylindrical jet will be completely evaporated once the jet has traveled a length $L_1$ \citep{bal06}. 
Taking the length of continuous H$\alpha$ emission in the inner part of the jet (not including clumps further along the flow axis so that we may assume that the filling factor $f$ is approximately unity) as $L_1$, we can derive an independent $\dot{M}$ estimate from the length of the H$\alpha$ jet and the well-constrained energetics in Carina. 
For a cylindrical jet, 
\begin{equation}
\dot{M} \approx \frac{L_1 f \mu m_H c_s}{2D} \left[ \frac{\alpha_B}{\pi r(d) L_{LyC} sin(\beta)} \right]^{-1/2} 
\end{equation}
where 
$f \approx 1$ is the filling factor for a cylinder losing mass from one side,  
$c_s \approx 11$ km s$^{-1}$ is the sound speed in photoionized plasma, and 
$\beta$ is the angle between the jet axis and the direction of the ionizing radiation, estimated from the H$\alpha$ image. 
For the eastern edge of the HH~901 flow ($\beta \approx 63^{\circ}$), this method gives $\dot{M} = 7.43 \times 10^{-6}$ M$_{\odot}$ yr$^{-1}$, more than an order of magnitude higher than the mass-loss rate derived from the H$\alpha$ emission measure ($1.84 \times 10^{-7}$ M$_{\odot}$ yr$^{-1}$). 
The disparity is even greater for the western limb of HH~901 ($\beta \approx 75^{\circ}$) where this method gives $\dot{M} = 3.77 \times 10^{-6}$ M$_{\odot}$ yr$^{-1}$, nearly two orders of magnitude higher than that derived from H$\alpha$ ($4.0 \times 10^{-8}$ M$_{\odot}$ yr$^{-1}$). 
Not surprisingly, new mass-loss rate estimates for both limbs of the jet are greater than $\dot{M}_{min}$. 

This method of estimating $\dot{M}$ yields $n_H$ larger than $n_{min}$, indicating that a slow-moving ionization front will propagate through the jet beam. 
Stratified emission in the western side of HH~901 seems to confirm this picture, with high surface brightness H$\alpha$ on the side of the jet closest to the source of ionizing radiation and brighter [Fe~II] emission further north, tucked $\sim 0.4$\arcsec\ behind the ionization front in the flow (see Section~\ref{sss:hh901} and Figure~\ref{fig:hh901_4view}). 

H$\alpha$-derived mass-loss rates are clearly underestimates for HH~901 and HH~902 (where the filling factor is somewhat more uncertain). 
While HH~666 and HH~1066 generally do not show the same stratified H$\alpha$ and [Fe~II] emission as HH~901 and HH~902, both have substantial [Fe~II] emission where there is little or no H$\alpha$ from the jet. 
This indicates that they also have a significant amount of shielded neutral H not traced by H$\alpha$. 
Both estimates of the mass-loss rate are, in principle, lower limits.
Clumpy [Fe~II] emission, even where the optical emission is relatively continuous, indicates additional mass in the jet not included in previous mass-loss rates.

\subsection{Comparison to Irradiated Jets in Orion}\label{ss:orion_jets}
Orion is the nearest region of high-mass star formation that also hosts a large collection of well-studied irradiated outflows \citep{rei98,bal01,bal06}. 
Jets in Orion tend to have a morphology that is very different from those in Carina. 
Many demonstrate C-shaped symmetry where the jet bends away from the source of ionizing radiation, due either to the bulk flow of material from the nebula (i.e. sidewinds) or to the rocket effect pushing on the neutral jet core \citep{bal01,bal06}. 
While there is some evidence that the rocket effect may bend HH~901, most of the jets in Carina do not, in general, show significant bending (4/39 of the larger sample do appear to be bent; all five of the HH jets in Orion reported by \citealt{bal01} are bent). 

If the rocket effect is the dominant mechanism for bending the jet, then we might expect the jets in Carina (subject to feedback from $\sim70$ O-type stars) might show a greater degree of bending. 
As discussed in \citet{bal06}, jet deflection by the rocket effect depends only on the mass of the jet and its evaporation rate, suggesting that these jets have remained remarkably straight because their high densities and masses provide the inertia needed to resist acceleration of the jet away from the ionizing source. 

HH~901 is only $\sim 1$ pc from Tr14, and thus is subject to the most direct and intense radiation of the jets in this sample. Its two limbs do not lie along the same axis (they differ by $\sim 12^{\circ}$), suggesting that it may be bent by the rocket effect. 
Thus, the observed deflection angle provides an independent way to estimate the density in HH~901. 
As a conservative estimate, we assume that both limbs of HH~901 started along the same axis, and have since each been deflected by $\sim 6^{\circ}$ and lie in the plane of the sky. 
The time for the western most edge of the continuous H$\alpha$ flow to reach its current location ($\sim 0.035$ pc from the pillar edge), assuming a jet velocity of $V = 200$ km s$^{-1}$, is $t = L_1 /V = 177$ yr. 
In that time, the western-most clump has been deflected $\sim 0.005$ pc, which gives a deflection velocity, $V_d \approx 28$ km s$^{-1}$. 

Following \citet{bal06}, this deflection velocity can be produced by the anisotropic photoablation of a neutral jet core, 
$V_d = c_s ln(m_0/m)$ where 
$m_0$ is the initial neutral H mass of a jet segment and 
$m$ is the remaining mass. 
Assuming mass loss is dominated by photoablation, the remaining mass is simply $m = m_0 - \dot{m} t$ where 
\begin{equation}
\dot{m} = f \pi \mu m_H c_s \frac{L_{LyC} \, r sin(\beta)}{4 \pi D^2 \alpha_B}
\end{equation} 
is the mass-loss rate per unit length in the jet due to photoevaporation. 

Solving for $m_0$, we find that the minimum mass of a spherical jet segment that survives $\sim 0.035$ pc from the pillar edge is $\sim 2 \times 10^{-5}$ M$_{\odot}$, which implies a density of at least $n_H = \frac{m_0}{\mu m_H \frac{4}{3} \pi r^3} = 3.2 \times 10^4$ cm$^{-3}$. 
This independent estimate of $n_H$ in HH~901 is similar to the minimum density derived in Section~\ref{ss:mdot_nd} and the critical density of the [Fe~II] lines.  
None of the other three jets show evidence for bending, suggesting that their densities are at least as high (although their photoablation rates are somewhat lower due to their greater distances from the source of ionizing radiation). 

One of the most striking differences between the irradiated outflows in the two regions is that the outflows in Orion are driven mostly by optically visible low-mass stars, while fewer than 20\% of HH jets in Carina have an exposed driving source. Only one of the four jets studied here has a driving source that can be identified in optical images (HH~666), although it is very faint, and can only be identified with HST. 

In Orion and Carina, the longest jets remain highly collimated over large distances. 
The Mach angle ($\mu = sin^{-1} \frac{c_s}{v_{jet}}$), and thus the opening angle ($\theta = 2\mu$), can remain small if the jet is highly supersonic. 
For the powerful jets in Carina, outflow speeds may exceed 200 km/s, although this is not necessarily the case for the jets in Orion \citep{bal01}. 
The sound speed in ionized material ($\approx 11$ km/s) is an order of magnitude larger than in neutral material, which suggests that the higher sound speed in the plasma on the irradiated side of the jet should lead to more jet spreading from an ionized jet than from a neutral jet. 
However, this is not observed. 
A partially ionized jet can maintain a greater degree of collimation if the ionized edge of the jet shields a significant cold, neutral core where the sound speed is lower. 
Given evidence for a significant amount of neutral material in the jets in Carina, this may help explain why the jets in Carina and Orion remain highly collimated. 
This is also a particularly appealing explanation for the high degree of collimation seen in the [Fe~II] jet in the eastern edge of HH~666 (feature O) where the H$\alpha$ emission is somewhat broader.  

Alternative explanations for the observed high degree of collimation include external pressure from the H~{\sc ii} region (although this is unlikely given the directional nature of the radiation pressure, especially near HH~901 and HH~902) and magnetic confinement. 
Theoretical models of magneto-hydrodynamic winds often argue for magnetic collimation of the flow \citep[e.g.][]{fen06,pud06} and large-scale, aligned magnetic fields have been observed in some sources \citep[e.g.][]{ray97,car10}. 
However, among the few protostellar sources for which magnetic fields have been measured, there are many examples of misaligned fields \citep{hul13}, making the role of magnetic fields in jet collimation unclear.

\subsection{Comparison to Other Massive Outflows}\label{ss:other_flows}
Most outflows driven by intermediate-mass protostars have been studied via their molecular emission \citep[e.g.][]{bel02,bel06,bel08,beu08,for09,fue09,tak12}. 
Higher M$_{\mathrm{ZAMS}}$ sources evolve faster along the pre-main sequence, so one expects that most examples of intermediate-mass protostellar outflows are still deeply embedded. 
A variety of models have been proposed to explain the broad spectrum of flow morphologies observed, although most have focused on outflows from low-mass protostars \citep{cab97,arc07}. 
Observed similarities between low- and relatively high-mass outflows suggest a similar production mechanism regardless of protostellar mass \citep{ric00}, although this has been disputed on the basis of observational biases \citep[particularly distance, see][]{rid01}. 

\begin{figure}%[t!]
\centering
$\begin{array}{c}
\includegraphics[trim=15mm 15mm 10mm 15mm,angle=-90,scale=0.25]{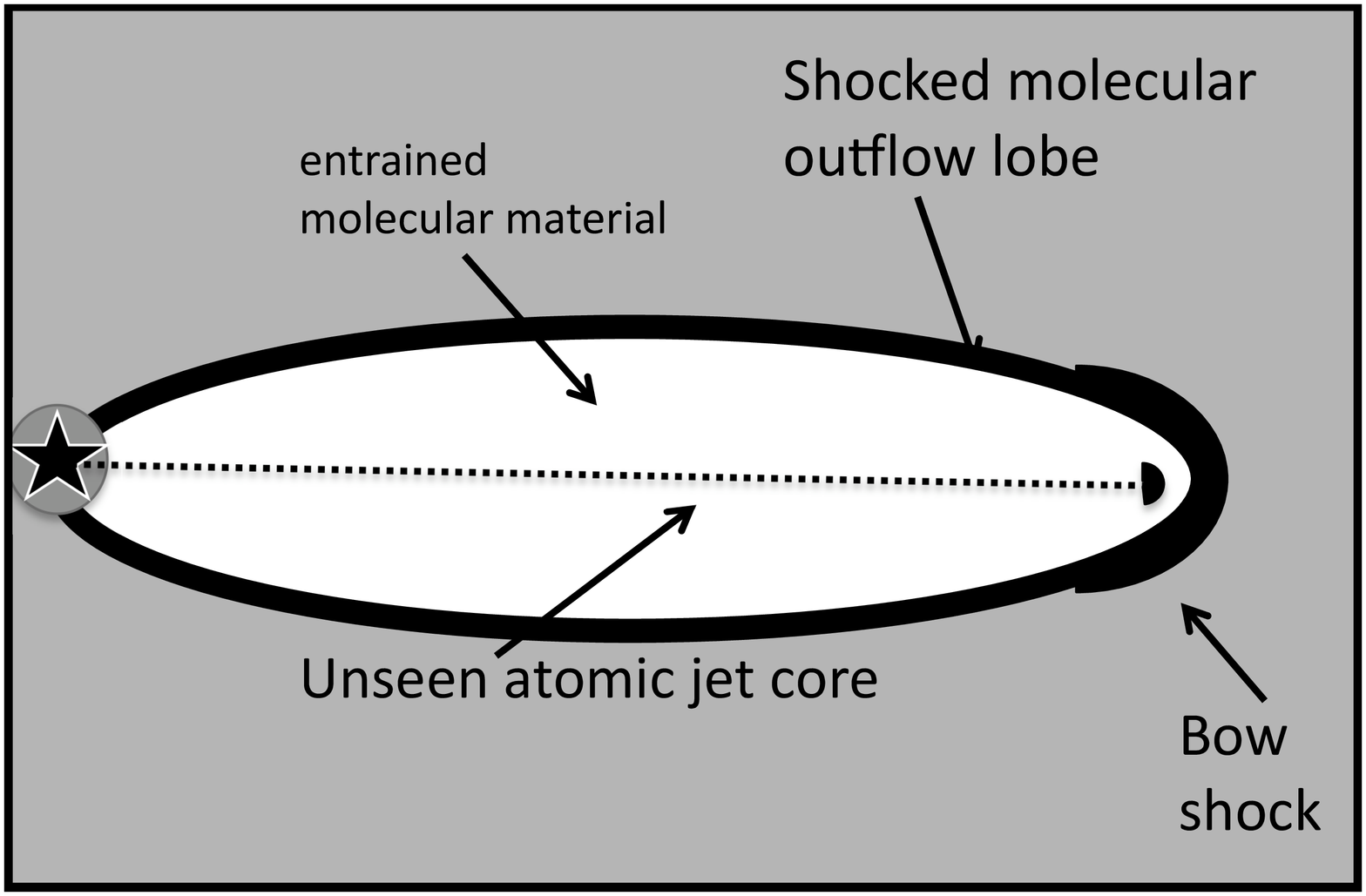} \\
\includegraphics[trim=15mm 15mm 15mm 15mm,angle=-90,scale=0.25]{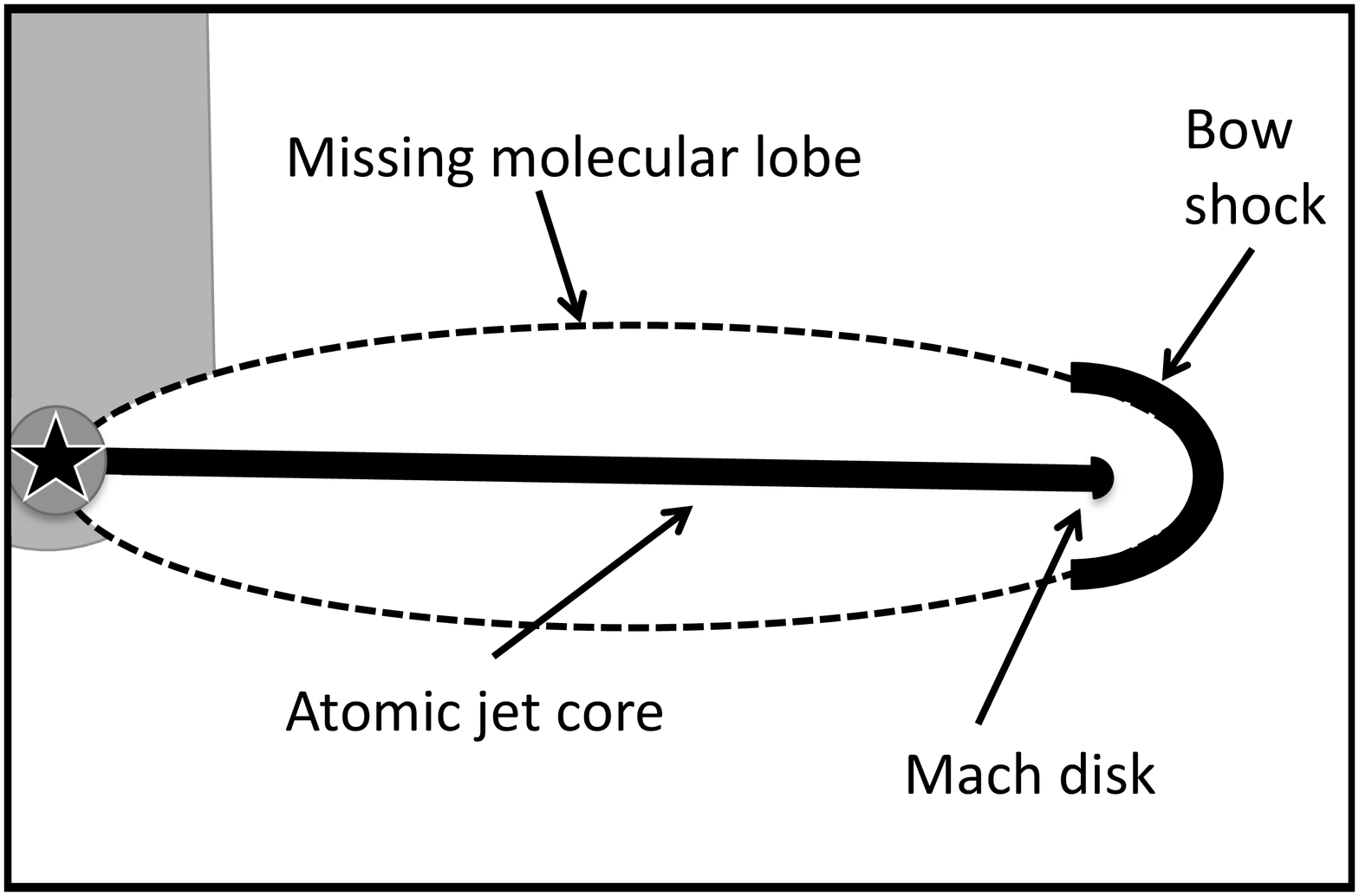} \\
\end{array}$
\caption{\textit{Top:} An illustration of the molecular jets often observed in embedded star-forming regions. Without an external ionizing source, only the molecules excited in walls of the outflow cavity trace the observed outflow while the atomic jet core remains invisible. 
Some of the jet core may be emit in the radio continuum if shocks in the flow lead to partial ionization. 
\textit{Bottom:} Outflows emerging into an H~{\sc ii} region will be bathed in UV radiation that will rapidly dissociate entrained molecules, leaving the irradiated atomic jet core to be gradually ionized and photo-evaporated outside the natal pillar. 
}\label{fig:evap_jet_sketch} 
\end{figure}

Thus, the Carina nebula is a particularly useful environment in which to study protostellar outflows. 
It provides a heterogeneous sample of outflows at a single distance that may be observed with uniform resolution. 
Carina houses a considerable number of intermediate-mass protostars near a large population of much more massive O-type stars (see Section~\ref{ss:environment}). 
If a jet-driven molecular outflow emerges into a giant H~{\sc ii} region like the Carina nebula, molecules in the flow will quickly be dissociated by the harsh UV radiation bathing the region (see Figure~\ref{fig:evap_jet_sketch}). 
Thus, only the irradiated atomic jet core will remain, allowing for direct study of the jet. 
This removes the need to infer jet properties from the emission of entrained molecules which will be complicated by variations in the environment into which the jet propagates. 
The lack of a wider envelope of entrained molecules may also explain why the irradiated jets in Carina and Orion are observed to be so highly collimated \citep{bal01,smi10a}. 

A nice example of a collimated jet and shocked envelope is HH~666 (Figure~\ref{fig:hh666_4view}). 
[Fe~II] emission traces the dense, highly collimated jet powering the HH~666 flow, following the jet into the molecular globule all the way back to the driving source. 
H$\alpha$ and [S~II] emission surround the [Fe~II] jet, particularly in feature HH~666~O, tracing a sheath around the jet that may be irradiated material in the walls of a cavity cleared by the jet (see Figure~\ref{fig:hh666_4view}d). 
Surprisingly, \citet{smi04} found no H$_2$ emission associated with the jet. 
Outside of the molecular globule, this is consistent with the rapid dissociation of molecules we would expect in the harsh UV environment created by the numerous O stars in the Carina nebula. 
Less clear is why there appears to be no H$_2$ emission from the jet inside the rich molecular environment of the pillar (although faint emission from the jet may be difficult to detect against the bright background). 

The HH jets in Carina comprise the largest sample of irradiated flows from intermediate-mass stars in a single region, and are among the most massive irradiated outflows known. 
Unlike molecular outflows from other intermediate-mass protostars, the HH jets in Carina can be studied directly (rather than via the molecular material they entrain) because they are not obscured by the extensive gas and dust that typically surrounds young star forming regions, and because the jet gas is externally illuminated. 
Few HH objects associated with more massive protostellar outflows have been found for this reason \citep[although see][]{mcg04}. 

Intermediate-mass stars forming alongside massive stars, like those in the Carina nebula, form in an environment that is profoundly changed by the H~{\sc ii} region carved by the most massive cluster members. 
As dust pillars harboring star formation are eroded by the advancing ionization front, accretion-driven outflows will emerge into the H~{\sc ii} region and will be irradiated while they are very young. 
Since most intermediate-mass stars will form alongside faster-evolving high-mass stars, these irradiated flows may be more representative of intermediate-mass protostellar outflows than their molecular counterparts.

\subsection{Environment}\label{ss:environment}
Within the Carina H~{\sc ii} region, the four outflows considered here are subjected to varying amounts of radiation and feedback depending on their distance and orientation relative to the ionizing radiation from nearby massive stars. 
One of the most striking differences is that [Fe~II] emission traces the jet back to a detectable \textit{Spitzer} point source in HH~666 and HH~1066, but for HH~901 and HH~902 there is no unambiguous IR protostar inside either pillar. 
The driving sources of both HH~901 and HH~902 could be extremely young and deeply embedded (Class 0), they could be obscured by an edge-on circumstellar disk, or they could emerge from extraordinarily dense globules that remain optically thick even into the mid-IR. 

Unlike HH~666 and HH~1066, HH~901 and HH~902 protrude into the hot, shocked stellar wind bubble expanding from the large O-star population in the centre of the nebula. 
Outside of this bubble, ionizing radiation drives a photoevaporative flow off of neutral pillars, creating a protective boundary layer that somewhat reduces the incident ionizing flux \citep{ber89}. 
This appears to be the case with HH~666 where extended streams of H$\alpha$ emission normal to the ionization front clearly trace the photoevaporative flow from the pillar (see Figure~\ref{fig:hh666_full}). 
Inside the stellar wind bubble, globules and pillars that survive the initial passage of the ionization front will be compressed, settling into a steady configuration where the pressure of the ionization front is balanced by the pressure of the neutral gas in the pillar \citep{ber90}. 
In this regime, extremely high densities are necessary to balance the pressure of the surroundings. 
Limits on the visual extinction of HH~901 imply a much larger column density ($N_H \gtrsim 2 \times 10^{22}$ cm$^{-2}$, see Section~\ref{sss:hh901}) than is measured for HH~666 ($N_H \approx 6 \times 10^{20}$ cm$^{-2}$), consistent with what \citet{pre12} found from their analysis of Herschel data. 
Together with the much smaller size of the HH~901 pillar compared to HH~666 (a factor of 10), the estimated volume density is nearly three orders of magnitude larger for HH~901 ($n_H \gtrsim 3 \times 10^{5}$ cm$^{-3}$ compared to $n_H \approx 4 \times 10^{2}$ cm$^{-3}$ for HH~666). 
This high density obscures the HH~901 driving source, even at $10-20$ \micron.

\subsection{Collimated Outflows from Intermediate-Mass Protostars}\label{ss:outflows_woot}
The intermediate-mass stars driving outflows in the Carina nebula add to a growing body of evidence that stars of all masses drive highly collimated outflows ($\theta \leq 10^{\circ}$) as part of their evolution \citep[e.g.][]{pal10,qiu11,zhu11}. 
While some observations \citep[e.g][]{wu04} and theory \citep[e.g.][]{vai11} suggest that more massive protostars drive less collimated flows ($\theta \geq 20^{\circ}$), we find that the protostellar jets observed in Carina are no less collimated than their low-mass counterparts (opening angles of a few degrees). 
\citet{mcg04} also find similarly small opening angles in their study of five HH jets driven by intermediate-mass stars.
Stronger radiation from more luminous, more massive protostars has been proposed as a physical mechanism to decrease outflow collimation \citep{beu05,vai11}. 
However, differences in collimation may also be an observational bias as outflows found to be poorly collimated at low resolution have been shown to be more collimated when observed at higher resolution \citep{beu02a}. 

This difference is particularly significant for outflows from intermediate- and high-mass stars, the majority of which have been observed with molecular emission lines in the millimeter (see Section~\ref{ss:other_flows}).
Molecular observations are blind to the atomic jet that may power the flow \citep[e.g.][]{dio09}, and only sample molecules in walls of a cavity or shock cone created by the jet. 
Shocks expand laterally with time, so an outflow observed only in molecular emission may appear broad while the invisible atomic jet remains highly collimated. 
A two-component jet system like this will look completely different in highly embedded regions \citep[e.g.][]{tor03,beu08,tak12} as compared to an irradiated environment like Carina (see Figure~\ref{fig:evap_jet_sketch}). 
Because the jets in Carina protrude into the H~{\sc ii} region, the UV radiation field quickly destroys molecules entrained in the flow outside the pillar, and illuminates the largely neutral atomic jet, revealing a highly collimated flow. 
The highly collimated jets in Carina suggest that the scenario of low-mass star formation can be ``scaled-up'' at least to $\approx 6-8$ \msun. 

\section{Conclusions}\label{s:conclusion}
We analyse high-resolution narrowband HST/WFC3-IR $\lambda 12657$ and $\lambda 16435$ [Fe~II] images of four of the most spectacular jets in the Carina nebula. 
HH~666, HH~901, HH~902, and HH~1066 all show bright IR [Fe~II] emission with $\lambda 16435/H\alpha$ ratios at least an order of magnitude higher than the ratio measured in nearby photoionized gas, revealing a significant amount of self-shielded, neutral material in the jet not traced by H$\alpha$. 
In both HH~901 and HH~902, H$\alpha$ is brightest along the side of the jet that faces the source of ionizing radiation (Tr14), while [Fe~II] emission peaks further away. 
This stratified emission structure reflects a slow-moving ionization front propagating through the width of the jet beam. 
The jet will be optically thick to Lyman continuum photons if the average atomic density, $n_H$, is sufficiently high, thus preventing the rapid ionization of the jet. 
For HH~901, we find $n_H \approx 1-3 \times n_{min}$, the minimum density to be optically thick to Lyman continuum radiation. 
In addition, we estimate the mass-loss rates of both limbs of HH~901 without assuming that the jet is fully ionized, and find values \textit{an order of magnitude larger} than those based on the H$\alpha$ emission measure. 
Bright [Fe~II] emission in all four jets suggests that the mass-loss rates derived from the H$\alpha$ emission measure vastly underestimate the true amount of mass lost in these jets, indicating that these jets are substantially more massive than previously thought. 
We emphasize that these constraints on the jet density come from the survival of Fe$^+$ against further ionization, and do not depend on the excitation mechanism that powers [Fe~II] emission. 
We do, however, note several morphological reasons to expect that the [Fe~II] emission is significantly influenced by FUV radiation. 

If the jet mass-loss rate is $\sim 1-10$\% of the accretion rate, more mass in these jets suggests that they are likely to be driven by intermediate-mass protostars ($2-8$ \msun) instead of low-mass stars ($< 2$ \msun). 
Model fits to the IR SEDs of the protostars driving HH~666 and HH~1066 ($6.3$ \msun\ and $2.8$ \msun, respectively) corroborate this interpretation, adding to a growing body of evidence that stars of all masses form by accretion.  
We find that the jets in Carina are no less collimated than their low-mass counterparts. 
The essential difference is that the optical and near-IR emission lines trace the neutral atomic jet core that would remain unseen if the jet were not irradiated in the H~{\sc ii} region. 
The harsh UV environment in Carina allows a unique view of these four outflows and provides new evidence that even relatively massive stars can drive collimated jets. 

[Fe~II] emission traces both HH~666 and HH~1066 back into the parent globule, connecting the jets to Spitzer-identified intermediate-mass protostars. 
No [Fe~II] emission associated with the jet is detected inside the pillars housing HH~901 and HH~902, nor is there any detection of point-like emission from the driving sources even at wavelengths as long as 24\micron. 
This may be a consequence of higher extinction, which is in turn created by the high pressure environment; HH~901 and HH~902 protrude into the hot stellar wind bubble created by Carina's many O stars. 
Large optical depths in the pillars may reflect high densities created as the pillar was compressed by the advancing hot bubble.

%%=============================================================================

\section*{Acknowledgments}
Support for this work was provided by NASA grant AR-12155 from the Space Telescope Science Institute. 
This work is based on observations made with the NASA/ESA Hubble Space Telescope, obtained from the Data Archive at the Space Telescope Science Institute, which is operated by the Association of Universities for Research in Astronomy, Inc., under NASA contract NAS 5-26555. These observations are associated with program GO/DD 12050.
We acknowledge the observing team for the HH~901 mosaic image:
Mutchler, M., Livio, M., Noll, K., Levay, Z., Frattare, L., Januszewski, W., Christian, C., Borders, T.

%%=============================================================================

\label{lastpage}

\end{document}